# A Subluminal Relativistic Magnetohydrodynamics Scheme with ADER-WENO Predictor and Multidimensional Riemann Solver-Based Corrector

By Dinshaw S. Balsara (dbalsara@nd.edu) and Jinho Kim (jkim46@nd.edu)

Physics Department, University of Notre Dame

**Abstract**

The relativistic magnetohydrodynamics (RMHD) set of equations has recently seen increased use in astrophysical computations. Even so, RMHD codes remain fragile. The reconstruction can sometimes yield superluminal velocities in certain parts of the mesh. The current generation of RMHD codes does not have any particularly good strategy for avoiding such an unphysical situation. In this paper we present a reconstruction strategy that overcomes this problem by making a single conservative to primitive transformation per cell followed by higher order WENO reconstruction on a carefully chosen set of primitives that guarantee subluminal reconstruction of the flow variables. For temporal evolution via a predictor step we also present second, third and fourth order accurate ADER methods that keep the velocity subluminal during the predictor step. The methods presented here are very general and should apply to other PDE systems where physical realizability is most easily asserted in the primitive variables.

The RMHD system also requires the magnetic field to be evolved in a divergence-free fashion. In the treatment of classical numerical MHD the analogous issue has seen much recent progress with the advent of multidimensional Riemann solvers. By developing multidimensional Riemann solvers for RMHD, we show that similar advances extend to RMHD. As a result, the face-centered magnetic fields can be evolved much more accurately using the edge-centered electric fields in the corrector step. Those edge-centered electric fields come from a multidimensional Riemann solver for RMHD which we present in this paper. The overall update results in a one-step, fully conservative scheme that is suited for AMR.

In this paper we also develop several new test problems for RMHD. We show that RMHD vortices can be designed that propagate on the computational mesh as self-preserving structures. These RMHD vortex test problems provide a means to do truly multidimensional accuracy testing for RMHD codes. Several other stringent test problems are presented. We show the importance of resolution in certain test problems. Our tests include a demonstration that RMHD vortices are stable when they interact with shocks.

**I) Introduction**

Just like non-relativistic magnetohydrodynamics (MHD), relativistic magnetohydrodynamics (RMHD) has seen a great deal of recent progress. Both MHD and RMHD are PDE systems with an involution constraint, where the magnetic field starts off divergence-free and stays so for all time. In a numerical code, keeping the magnetic field divergence-free requires the use of a staggered, Yee-type mesh. Likewise, both systems are non-convex with the result that their numerical treatment results in codes that tend to be more brittle than a traditional hydrodynamics code. Numerical RMHD suffers from the further deficiency that reconstructed velocities can become superluminal at certain locations of the mesh, thus making RMHD codes even more fragile than MHD codes. Several advances have been made in numerical RMHD (e.g. Aloy *et al*., [3], Komissarov [50], Balsara [7], Del Zanna *et al*. [31], Gammie *et al*. [42], Komissarov [51], Ryu *et al*. [62], Mignone & Bodo [56], Honkkila & Janhunen [47], Del Zanna *et al*. [32], Tchekhovskoy *et al*. [64], Mignone, Ugliano and Bodo [57], Anton *et al*. [4], McKinney *et al*. [55], Kim & Balsara [49], Etienne *et al*. [39]). In many of these papers, several very sophisticated one-dimensional Riemann solvers have been designed and many very nice higher order reconstruction strategies have been discussed. In other words, this is the traditional progress that needs to be made for producing higher order Godunov schemes for RMHD. The fact that such progress has been made is a very good thing because it sets the stage for the next round of progress reported here.

Despite all the excellent progress, RMHD codes remain brittle. Ensuring positivity of density or pressure can be a challenge. Fortunately, work done in ensuring the positivity of density and pressure in non-relativistic MHD (Balsara [13]) can be directly transcribed to RMHD. Consequently, ensuring the positivity of density and pressure in RMHD simulations is not such a big challenge any more. However, RMHD codes suffer from a further deficiency. They can frequently produce reconstructed velocities that might be superluminal in certain places on the mesh. This is a bigger difficulty because once the flow becomes superluminal, it is impossible for the RMHD code to recover. This problem is exacerbated by the fact that some RMHD calculations require very high Lorentz factors, and an increasing Lorentz factor makes it more likely that superluminal flow might develop somewhere on the computational mesh. The *first goal* of this paper is to show that there is a way to reformulate the reconstruction problem that entirely avoids situations where the reconstructed velocities can become superluminal. Our method is very efficient because it involves a single conservative to primitive transcription per zone along with a single WENO (Weighted Essentially Non-Oscillatory) reconstruction step. In other words, our method is as fundamentally efficient as reconstruction on the conserved variables while guaranteeing a sub-luminal reconstruction of the velocities. A variant of this reconstruction algorithm, as it applies to two-fluid relativistic electrodynamics, has already been presented in Balsara *et al*. [22]. Our reconstruction method is very general and should apply to high order treatment of other recondite PDE systems where positivity or other physical constraints are more easily imposed on the primitive variables.

Accurate spatial reconstruction is only useful if it is matched with a suitably high order accurate temporal evolution. Typically, one wishes to have temporal evolution with the same order of accuracy as the spatial evolution. While higher order multistage Runge-Kutta (RK) schemes can be designed for RMHD, we argue that it provides an inferior choice. The temporal evolution of the velocity in the predictor step of a higher order scheme should also keep the velocity subluminal by design. This *second goal* of keeping the velocity subluminal during the predictor step is most easily realized by ADER (Arbitrary DERivatives in space and time) schemes as shown in a subsequent section. We therefore present ADER schemes at second, third and fourth orders of temporal accuracy that preserve the subluminal evolution of the velocity while doing so at a very low computational cost. In some sense, the ADER schemes presented here are optimized for RMHD and perhaps other recondite PDE systems that might share the same numerical challenges as RMHD. (It is good to keep in mind that RMHD is, in some sense, the lowest order approximation for a relativistic plasma. Consequently, relativistic plasma physics simulations might have other, more difficult, equation sets. To wit, consider the equations that describe a coupled, magnetized electron-positron plasma that has recently been dealt with in Balsara *et al*. [22].) As in the previous paragraph, we mention that the idea of evolving the primitive variables with an ADER scheme is very general and should apply to high order treatment of other recondite PDE systems where positivity or other physical constraints need to be retained in the predictor step. For those who prefer the simplicity of SSP-RK methods, it is valuable to point out that in Balsara *et al*. [22] we also presented a high order strategy for treating stiff source terms with IMEX-RK methods that is particularly simple.

There has also been a lot of recent work to put divergence-free MHD codes on a firm footing. This work is based on the realization that any divergence-free formulation of MHD requires the use of a Yee-type mesh where the magnetic field components are collocated at face centers and the electric fields are collocated at edge centers. As shown in Balsara & Spicer [5], the electric field is just given by suitable components of the flux when the MHD system is written in flux-conservative form. This insight extends to RMHD. The problem though is that the fluxes have to be evaluated multidimensionally at the edges of the mesh in order to evaluate the electric fields that reside there. This necessitates the development of multidimensional Riemann solvers that can operate at the edges of a mesh. For hydrodynamics and non-relativistic MHD a substantial body of work on multidimensional Riemann solvers already exists (Balsara [12, 14, 17, 20], Balsara, Dumbser & Abgrall [16], Vides, Nkonga & Audit [69], Balsara & Dumbser [18, 19], Balsara *et al*. [21], Balsara *et al*. [23]). These papers have shown that multidimensional Riemann solvers completely eliminate any need to double the dissipation in the evaluation of the electric field. Balsara [13, 14, 17] showed that by enabling the magnetic field to be evolved more accurately on the mesh, these multidimensional Riemann solvers permit us to treat very low-$\beta$ plasma flows. Such low-$\beta$ plasma flows, which have very strong magnetic fields and very low thermal pressures, are very common in non-relativistic MHD. Analogous flows with very high magnetization are also very common in RMHD. Consequently, there is a need to extend the previously-reported advances to RMHD. The *third goal* of this paper is to show that multidimensional Riemann solvers can be formulated for RMHD. Balsara [14, 17] showed that low dissipation variants of such Riemann solvers can be designed which preserve sub-structure in the resolved state. Self-similarity is crucially important in the development of multidimensional Riemann solvers. This has prompted the name of MuSIC Riemann solvers, where MuSIC stands for "Multidimensional, Self-similar, strongly-Interacting, Consistent". For a video introduction to multidimensional Riemann solvers please see: http://www.nd.edu/~dbalsara/Numerical-PDE-Course. In this paper we extend MuSIC Riemann solvers to RMHD. Specifically, we draw on the formulation from Balsara [17] and Balsara *et al*. [23] to catalogue multidimensional Riemann solvers for RMHD that preserve the substructure in the resolved state.

The WENO reconstruction and ADER predictor step have to be complemented with a corrector step. This corrector step entails the evaluation of the facially-averaged numerical fluxes for the conserved variables as well as the edge-averaged values of the electric fields. This is the corrector step of our one-step update algorithm. Such one-step methods are known to be very valuable for AMR. In prior paragraphs we have mentioned the value of subluminal spatial reconstruction via WENO and the usefulness of a subluminal predictor step via ADER. It is also important to mention that positivity considerations are also very valuable in the corrector step. Unfortunately, positivity proofs are only available when certain classes of Riemann solvers are used and, moreover, they only apply to first order schemes. Einfeldt [37] and Einfeldt *et al*. [38] were only able to demonstrate positivity for an HLL/HLLEM class of Riemann solvers applied to one-dimensional first order schemes for Euler flow. For HLLC-type Riemann solvers, again such proofs exist for one-dimensional Euler flow and one-dimensional MHD (Toro, Spruce & Speares [68], Batten *et al*. [25], Gurski [46]). For Roe-type Riemann solvers (Roe [61]) or Osher-Solomon type Riemann solvers (Osher and Solomon [59], Dumbser and Toro [35], Zanotti and Dumbser [71]), such positivity proofs do not currently exist. We see, therefore, that proofs that guarantee physical realizability are very scanty. When they exist, they apply to only a few PDE systems, and for one-dimensional flow based on first order schemes. For RMHD, there is no work on Riemann solvers that are provably subluminal. However, the prior discussion shows that methods that are based on HLL/HLLEM Riemann solvers or HLLC Riemann solvers have a greater likelihood of success. For that reason, all one-dimensional Riemann solvers used here are either based on relativistic HLLC Riemann solvers (Mignone & Bodo [56], Kim & Balsara [49]) or relativistic HLLEM Riemann solvers (Dumbser & Balsara [36], Balsara *et al*. [23]). The multidimensional Riemann solvers used here are also based on similar building blocks (Balsara [17], Balsara *et al*. [23]). For multidimensional problems with higher order of accuracy, the positivity of density and pressure and the preservation of a subluminal velocity cannot be guaranteed for RMHD. However, our corrector step is built on the best possible building blocks that are known to preserve physical realizability for Euler flow in the literature.

For the sake of completeness we mention here that very high order reconstruction of divergence-free MHD flows also requires drawing on the work of Balsara [8, 9, 10]. That work carries over directly to RMHD. Therefore, in this paper we restrict our focus to second, third and fourth order schemes for RMHD and show that within this context the three goals from the previous paragraphs can indeed be realized.

For the sake of completeness, and also because we want to put this work in context, we mention that there has been prior effort at designing multidimensional Riemann solvers. Abgrall [1, 2] designed the first of the genuinely multidimensional Riemann solvers for CFD. Further advances were also reported (Fey [40, 41], Gilquin, Laurens & Rosier [45], Brio, Zakharian & Webb [28]). Most of these above-mentioned genuinely multidimensional Riemann solvers did not see much use because they were difficult to implement. Wendroff [70] formulated a two-dimensional HLL Riemann solver, but his method was also not easy to implement and did not provide the numerical fluxes in closed-form. The recent advances on multidimensional Riemann solvers by the first author have, therefore, been crucial in putting MHD and RMHD on a conceptually sound footing.

The plan of this paper is as follows. In Section II we discuss the choice of reconstruction variables in RMHD, showing that there is an optimal choice that avoids the formation of superluminal velocities even in multidimensional flow. Once the spatial reconstruction is done, one wishes to evolve the solution "in-the-small" within each zone. This is the predictor step. In Section III we describe how this predictor step can be carried out in a way which keeps the velocity subluminal. In Section IV we present the formulation of the multidimensional Riemann solver for RMHD. Section V presents several novel accuracy tests that can be used to demonstrate the accuracy of RMHD codes. Section VI presents several stringent test problems. Section VII presents conclusions.

**II) Spatial Reconstruction: The Choice of Reconstruction Variables**

Let us begin by posing the question, "Which variables are the best variables for carrying out reconstruction in relativistic magnetohydrodynamics (RMHD)?" This is a very important question for relativistic hydrodynamics and for RMHD. It is an important question because the applications of interest require the RMHD codes to solve multidimensional problems with Lorentz factors up to a thousand! Unfortunately, most RMHD codes usually produce superluminal motions once Lorenz factors of ~10 are exceeded. Once the code goes superluminal in even one zone, there is no good corrective action (the root solvers fail) and a code crash often results. For relativistic hydrodynamics the issues are slightly simpler because it is quite easy to go from conserved variables to primitive variables and vice versa. For RMHD, the transcription is not quite so simple. Some of our recent work has focused on restoring fidelity to non-relativistic hydrodynamics and MHD calculations (Balsara [13]). In that same spirit, we focus on the process of restoring fidelity to RMHD calculations in this paper.

For the sake of completeness, let us first document the RMHD equations. The RMHD equations can be written in three dimensions as $\partial_t \mathbf{U} + \partial_x \mathbf{F} + \partial_y \mathbf{G} + \partial_z \mathbf{H} = 0$. This is written more explicitly as

$$\frac{\partial}{\partial t} \begin{pmatrix} \rho \gamma \\ m_x \\ m_y \\ m_z \\ \varepsilon \\ B_x \\ B_y \\ B_z \end{pmatrix} + \frac{\partial}{\partial x} \begin{pmatrix} \rho \gamma v_x \\ m_x v_x + P_{Tot} - B_x b_x/\gamma \\ m_y v_x - B_x b_y/\gamma \\ m_z v_x - B_x b_z/\gamma \\ m_x \\ 0 \\ B_y v_x - v_y B_x \\ B_z v_x - v_z B_x \end{pmatrix}$$

$$+ \frac{\partial}{\partial y} \begin{pmatrix} \rho \gamma v_y \\ m_x v_y - B_y b_x/\gamma \\ m_y v_y + P_{Tot} - B_y b_y/\gamma \\ m_z v_y - B_y b_z/\gamma \\ m_y \\ B_x v_y - v_x B_y \\ 0 \\ B_z v_y - v_z B_y \end{pmatrix} + \frac{\partial}{\partial z} \begin{pmatrix} \rho \gamma v_z \\ m_x v_z - B_z b_x/\gamma \\ m_y v_z - B_z b_y/\gamma \\ m_z v_z + P_{Tot} - B_z b_z/\gamma \\ m_z \\ B_x v_z - v_x B_z \\ B_y v_z - v_y B_z \\ 0 \end{pmatrix} = 0$$

(2.1)

Here $\rho$ is the density, $\gamma$ is the Lorentz factor, $\{\varepsilon, m_x, m_y, m_z\}$ form the four-momentum density, $v_x, v_y, v_z$ are the three-velocities, $P_g$ is the gas thermal pressure, $P_{Tot}$ is the total pressure and $B_x, B_y, B_z$ are the components of the magnetic field. The speed of light is taken to be unity so that the Lorentz factor is defined by $\gamma = 1/\sqrt{1-\mathbf{v}^2}$. The momentum and energy are related to the other variables by

$$m_i = (\rho h \gamma^2 + B^2)v_i - (v \cdot B)B_i \quad ; \quad \varepsilon = \rho h \gamma^2 - P_g + \frac{B^2}{2} + \frac{v^2 B^2 - (v \cdot B)^2}{2} \tag{2.2}$$

where $h = 1 + \Gamma P/((\Gamma - 1)\rho)$ is the specific enthalpy and $\Gamma$ is the polytropic index. The covariant magnetic field ($b^\mu$) and total pressure ($P_{Tot}$) are defined as

$$b_\mu = \gamma \left( v \cdot B, \frac{B_i}{\gamma^2} + (v \cdot B) v_i \right), \quad P_{Tot} = P_g + \frac{|b|^2}{2}, \quad \text{with} \quad |b|^2 = b^\mu b_\mu = |B|^2/\gamma^2 + (v \cdot B)^2. \tag{2.3}$$

In eqns. (2.2) and (2.3), Latin indices range from 1 to 3, whereas Greek indices range from 0 to 3; in keeping with relativistic conventions. Del Zanna *et al.* [32] provide an efficient transcription strategy from conserved to primitive variables.

One option is always to reconstruct in the conserved variables "**U**". Here "**U**" is the eight component vector of conserved variables. The reconstruction can be carried out componentwise or in characteristic variables. The problem with reconstructing on the conserved variables is that we eventually need to go back to the primitive variables and the transcription is non-trivial and computationally expensive for RMHD. Furthermore, just because one has a vector of conserved variables defined at any point in the zone (via piecewise linear reconstruction) does not ensure that the same vector is physical over the entire zone. It may have negative density or pressure (which is easy to fix by using the work of Balsara [13]). But, even more damaging, it may have superluminal motion. Once a zone in the mesh becomes superluminal, the calculation is invariably corrupted and the code crashes.

For the above-mentioned reasons, people sometimes prefer to reconstruct the vector of primitive variables $V = (\rho, v_x, v_y, v_z, P_g, B_x, B_y, B_z)^T$. When this is done for non-relativistic MHD, the process works well and can even be extended to include third order schemes. For RMHD, the reconstruction of the primitive variables has its deficiencies. Let us explain those deficiencies next. Imagine one zone with $v_x = 0.999$ (i.e., $\gamma = 22.366$) and a neighboring zone with $v_x = 0.9999$ (i.e., $\gamma = 70.712$). If we reconstruct the x-velocity, the difference between the zones would only yield an undivided difference $\Delta v_x = 0.0009$. This is a very small difference and prone to error accumulation. The gradient in the variables would then be much smaller than the variable being reconstructed. It can result in loss of fidelity, since the smallest increase in $\Delta v_x$ can drive the flow superluminal. It is precisely because velocities in high speed relativistic flow all tend to bunch up at ~1 that the velocity becomes a bad variable in which to carry out the reconstruction.

Indeed, there is a better way. The better way is to reconstruct in the special vector of primitive variables given by $V = \left( \rho, \gamma v_x, \gamma v_y, \gamma v_z, P_g, B_x, B_y, B_z \right)^T$. In other words, we are reconstructing the spatial components of the four-velocity. The inclusion of the Lorenz factor causes the variations from one zone to the next to become visible. Now the same two zones in the above example have $\gamma v_x = 22.344$ and $\gamma v_x = 70.704$. Clearly, we can reconstruct the slopes for $(\gamma v_x)$, $(\gamma v_y)$ and $(\gamma v_z)$; we can still hope to retain a significant variation in the last three components of the four-velocity.

Once the reconstruction is done, we can obtain $\left( \gamma v_x, \gamma v_y, \gamma v_z \right)^T$ anywhere within the zone. We now show that we can easily retrieve the three-velocity with just a few float point operations. We define the $\vartheta$ variable as follows:

$$\vartheta \equiv \gamma^2 v_x^2 + \gamma^2 v_y^2 + \gamma^2 v_z^2 = \frac{\vec{v}^2}{1-\vec{v}^2} \quad \Rightarrow \quad \vec{v}^2 = \frac{\vartheta}{1+\vartheta} \tag{2.4}$$

Notice, therefore, that with $\vartheta \geq 0$, we are assured that the three-velocity remains less than unity! I.e., if we use the present approach to retrieving the three-velocity then we can guarantee that the three-velocity will remain subluminal. We also find

$$\gamma = \sqrt{1+\vartheta} \tag{2.5}$$

which enables us to obtain the Lorenz factor at any point within a zone. Again, with $\vartheta \geq 0$, it is mathematically provable that we will always have $\gamma \geq 1$. In the non-relativistic limit, $\gamma \rightarrow 1$, these variables seamlessly yield the conventional strategy for reconstructing on the primitive variables. Thus our choice of reconstruction variable graciously reduces to the correct non-relativistic limit for hydrodynamics and MHD. Note too that our strategy is equally valid for structured and unstructured meshes.

On a historical note, it is worth pointing out that reconstruction on the spatial components of the four-velocity has also been suggested by Komissarov [50], Aloy et al. [3] and Balsara [7]. Their suggestion only yields second order schemes. We now show how higher order reconstruction can be carried out on the primitive variables to yield higher order schemes for RMHD that retain subluminal flow velocities.

Our higher order reconstruction strategy draws its inspiration from eqns. (12) and (16) from McCorquodale and Colella [54]. Let $\bar{\mathbf{U}}_{i,j,k}$ denote the zone-averaged vector of conserved variables in zone $(i,j,k)$. We wish to extract $\mathbf{U}_{i,j,k}$ which gives us the value of the vector of conserved variables exactly at the center of that zone. We will soon show that it is possible to define a flattener function, $\eta_{i,j,k}$, within each zone such that $\eta_{i,j,k} = 0$ for smooth flow and $\eta_{i,j,k} \to 1$ for non-smooth flow. Let function $\phi_{i,j,k} = 1 - \eta_{i,j,k}$ within that zone such that $\phi_{i,j,k} = 1$ for smooth flow and $\phi_{i,j,k} \to 0$ when the flow becomes increasingly non-smooth. The description of such flattener functions for various PDE systems is given in Colella and Woodward [29] and Balsara [13]. If the flow is smooth, i.e. if $\phi_{i,j,k} = 1$, fourth order accuracy can be ensured. Following McCorquodale and Colella [54], in the fourth order limit we can assert

$$\mathbf{U}_{i,j,k} = \bar{\mathbf{U}}_{i,j,k} - \frac{1}{24} \phi_{i,j,k} \left[ \left( \bar{\mathbf{U}}_{i+1,j,k} - 2\bar{\mathbf{U}}_{i,j,k} + \bar{\mathbf{U}}_{i-1,j,k} \right) + \left( \bar{\mathbf{U}}_{i,j+1,k} - 2\bar{\mathbf{U}}_{i,j,k} + \bar{\mathbf{U}}_{i,j-1,k} \right) + \left( \bar{\mathbf{U}}_{i,j,k+1} - 2\bar{\mathbf{U}}_{i,j,k} + \bar{\mathbf{U}}_{i,j,k-1} \right) \right]$$

(2.6)

In other words, when the flow is sufficiently smooth, eqn. (2.6) will give us a fourth order accurate value for $\mathbf{U}_{i,j,k}$, the vector of conserved variables which is defined pointwise at the zone center. Using standard root solver techniques that are available in the literature (Del Zanna *et al.* [32]), we obtain $\mathbf{V}_{i,j,k}$, the vector of primitive variables defined pointwise at the zone center, from the vector $\mathbf{U}_{i,j,k}$. Once $\mathbf{V}_{i,j,k}$ is obtained, we can obtain the zone-averaged vector of primitive variables $\bar{\mathbf{V}}_{i,j,k}$ by using the inverse procedure to eqn. (2.6). We get

$$\bar{\mathbf{V}}_{i,j,k} = \mathbf{V}_{i,j,k} + \frac{1}{24} \phi_{i,j,k} \left[ \left( \mathbf{V}_{i+1,j,k} - 2\mathbf{V}_{i,j,k} + \mathbf{V}_{i-1,j,k} \right) + \left( \mathbf{V}_{i,j+1,k} - 2\mathbf{V}_{i,j,k} + \mathbf{V}_{i,j-1,k} \right) + \left( \mathbf{V}_{i,j,k+1} - 2\mathbf{V}_{i,j,k} + \mathbf{V}_{i,j,k-1} \right) \right]$$

(2.7)

Now that we have the zone-averaged vector of primitive variables $\bar{\mathbf{V}}_{i,j,k}$ in all the zones, we can use that information to carry out multidimensional finite volume WENO or PPM reconstruction within the zone with up to fourth order of accuracy (Colella and Woodward [29], Jiang and Shu [48], Balsara and Shu [6], Balsara *et al*. [11], Colella and Sekora [30], McCorquodale and Colella [54]). Methods from Balsara [13] can now be applied to the reconstruction to ensure that the density and pressure are positive. The special choice of primitive variables from this paper ensures that the velocity remains sub-luminal. When the reconstruction is done in primitive variables, ensuring the positivity of the pressure is just as easy as ensuring the positivity of the density; and we know from eqns. (6) to (8) of Balsara [13] that enforcing positivity of density is indeed very easy.

For the sake of completeness, it is also helpful to explicitly define the flattener function. Such functions have been defined in Colella and Woodward [29] and Balsara [13] and rely on comparing the divergence of the velocity to a characteristic speed in the problem. For RMHD, we advocate using the undivided divergence of the spatial part of the four-velocity. While this measure is not Lorentz invariant, the inclusion of the Lorentz factor in the divergence ensures that small changes in speeds that are close to the speed of light can be registered by the flattener. Besides, please realize that the introduction of a computational mesh implies that we want the computed answers to be salient in a particular frame of reference. Our flattener is, therefore, defined in a frame-dependent fashion as

$$\eta_{i,j,k} = \frac{\left|(\nabla \cdot \gamma \mathbf{v})_{i,j,k}\right| \max(\Delta x, \Delta y, \Delta z)}{\kappa_1 \langle c_{i,j,k} \rangle_{\min}} - 1 \quad ; \quad c_{i,j,k} = \sqrt{\frac{\Gamma(\Gamma-1)\, P_{i,j,k}}{(\Gamma-1)\, \rho_{i,j,k} + \Gamma\, P_{i,j,k}}} \quad ; \qquad (2.8)$$

$$\eta_{i,j,k} = \min\left[\, 1,\ \max\left[\, 0,\ \eta_{i,j,k}\,\right]\,\right]$$

Here $c_{i,j,k}$ is a characteristic speed in the flow, preferably one that excludes the mean velocity of the flow. As a result, we choose $c_{i,j,k}$ to be the local sound speed in a relativistic fluid. Since shocks can propagate by one zone in any timestep, the flattener needs to be aware of the local sound speed in the 26 (three-dimensional) zones that surround a given zone. Consequently, $\langle c_{i,j,k} \rangle_{\min}$ is the minimum of the sound speed in the current zone and its immediate neighbors. The mesh size in the *x*-direction is given by $\Delta x$, and so on. In eqn. (2.8), $\kappa_1$ is taken to be a small number and is usually set to 0.2 to 0.4. We, therefore, see that the flattener $\eta_{i,j,k} = 0$ when the flow is smooth and it goes to $\eta_{i,j,k} = 1$ in a continuous fashion when strong shocks or rarefactions are present. To enhance the stabilization in zones that are about to be run over by a strong shock, we focus on the pressure profile in each direction. Thus, for the *x*-direction we have

$$\begin{aligned} if\ \left(P_{i+1,j,k} > \kappa_2 P_{i,j,k}\right) then\ \ \eta_{i,j,k} &\rightarrow \max\left(\eta_{i,j,k}, \eta_{i+1,j,k}\right) \\ if\ \left(P_{i-1,j,k} > \kappa_2 P_{i,j,k}\right) then\ \ \eta_{i,j,k} &\rightarrow \max\left(\eta_{i,j,k}, \eta_{i-1,j,k}\right) \end{aligned} \quad (2.9)$$

Here $\kappa_2 \sim 1.5$. If the flattener indicates the presence of a very strong shock in a zone, it may even be acceptable to resort to a TVD reconstruction in that zone; this was how Colella and Woodward [29] intended the flattener to be used. The description of the flattener completes the description of the subluminal spatial reconstruction scheme for RMHD at high orders.

The careful reader should also be cautioned that recently Zanotti and Dumbser [71] have suggested that it is acceptable to carry out the higher order reconstruction on the primitive variables given by $V = \left(\rho, v_x, v_y, v_z, P_g, B_x, B_y, B_z\right)^T$. This choice of primitive variables has no special advantage with respect to retaining subluminal velocities in RMHD. This point was also made in our talks on subluminal RMHD at the HONOM meeting in Trento on March 2015 and the Astronum meeting in Avignon on June 2015. While Zanotti and Dumbser attended our presentations, we are surprised that our point went unappreciated. It should further be noted that the reconstruction in Zanotti and Dumbser [71] requires two reconstruction steps, the second of which is non-standard. The method presented here requires only one very standard (PPM or WENO) finite volume reconstruction step. This enables implementers to benefit from the pre-existing reconstruction methods that are already present in their codes.

**III) The Predictor Step: ADER Timestepping in Primitive Variables**

The spatial reconstruction described in Section II was such that the velocity will remain subluminal by design throughout the zone in which this reconstruction is carried out. Temporal accuracy is endowed to these schemes by using either a semi-discrete in time Runge-Kutta scheme or by using a continuous in time predictor-corrector approach. Within the context of multistage Runge-Kutta schemes, there is no obvious way of ensuring that the velocity remains sub-luminal during all of the internal stages of the scheme. For that reason, we prefer predictor-corrector type approaches. In the design of higher order schemes of predictor-corrector type, the predictor step uses the spatial variation within a zone to evolve the PDE system in time to the appropriately high order. It is highly desirable to keep the velocity sub-luminal during this process. This can be achieved if the four-velocity variables are evolved in time. In other words, despite the overall PDE system for RMHD being in conservation form, it pays to carry out the predictor step in our special choice of primitive variables.

For RMHD, the transcription from conserved to primitive variables is also prohibitively expensive. Keeping the time evolution in primitive form keeps the predictor step relatively inexpensive. But please realize that the predictor step cannot be made very inexpensive because obtaining the governing equations in primitive form still involves the inversion of a matrix, as we will show in time. For RMHD, this inversion cannot be carried out analytically; so it has to be done numerically.

ADER schemes have established themselves as favored methods for instituting such a predictor step for the numerical solution of PDE systems. We therefore work within the context of ADER schemes. Early formulations (Titarev & Toro [65, 66], Toro & Titarev [67]), which were based on Riemann solvers, have now given way to Galerkin-like formulations (Dumbser *et al*. [33, 34], Balsara *et al*. [11, 15]). We therefore prefer the latter variant of ADER schemes. Operationally, these methods consist of establishing a suitable set of nodal points in space-time. At each of those points, the PDE solution is enforced. This is then used to drive the space-time representation of the PDE system within the zone to convergence. This is done via iteration, but the iterative process is strongly contractive so that convergence is easily obtained within a few iterations. Such an iterative predictor step can be instituted in conserved or primitive variables. For RMHD, sub-luminal velocities can only be ensured by design if the four-velocity itself is evolved in time. For that reason, we prefer to formulate ADER schemes in primitive variables. Recall too, that enforcing the PDE at each nodal point involves the inversion of a small 8×8 matrix at that nodal point. This can become an expensive process especially if the inversion is required at too many nodal points. For that reason, the ADER schemes that we formulate in primitive variables are designed to minimize the number of matrix inversions.

In the next three sub-sections we explicitly catalogue ADER schemes at second, third and fourth orders. The second order scheme is very easy to understand and reduces to a classical Lax-Wendroff procedure. To aid comprehension, we therefore present it as a classical Lax-Wendroff predictor step. This might also be useful to implementers who want to stop short at second order and do not want to invest time in understanding the intricacies of ADER schemes. At third and fourth orders, the formulation will be presented in the true spirit of an ADER scheme.

The fact that our ADER scheme is imposed on the primitive variables in no way affects the overall order of accuracy of the scheme.

### III.a) Second Order ADER Scheme; a.k.a. Lax-Wendroff type Predictor

The favored set of primitive variables in this paper is given by the vector $V = \left( \rho, \gamma v_x, \gamma v_y, \gamma v_z, P, B_x, B_y, B_z \right)^T$. Any second order accurate spatial reconstruction applied within a given zone will yield three further undivided difference vectors, $\Delta_x V$, $\Delta_y V$ and $\Delta_z V$ within that zone. Let $\Delta x$, $\Delta y$ and $\Delta z$ be the zone size in the x-, y- and z-directions and let $\Delta t$ be the current timestep. The timestep is intended to carry the solution from an initial time $t^n$ to a final time $t^{n+1} = t^n + \Delta t$. Using the variables mentioned in the previous few sentences, we wish to find $\Delta_t V$, the undivided difference in time that gives us the time-evolution that is consistent with the governing equations. Eqn. (2.1) can be formally written as $\partial_t \mathbf{U} + \partial_x \mathbf{F} + \partial_y \mathbf{G} + \partial_z \mathbf{H} = 0$ and it can be written in primitive variables as

$$\frac{\partial V}{\partial t} + A \frac{\partial V}{\partial x} + B \frac{\partial V}{\partial y} + C \frac{\partial V}{\partial z} = 0$$

$$\text{with } A \equiv \left( \frac{\partial \mathbf{U}}{\partial V} \right)^{-1} \left( \frac{\partial \mathbf{F}}{\partial V} \right) \; ; \; B \equiv \left( \frac{\partial \mathbf{U}}{\partial V} \right)^{-1} \left( \frac{\partial \mathbf{G}}{\partial V} \right) \; ; \; C \equiv \left( \frac{\partial \mathbf{U}}{\partial V} \right)^{-1} \left( \frac{\partial \mathbf{H}}{\partial V} \right)$$

(3.1)

The matrices $\partial \mathbf{U}/\partial V$, $\partial \mathbf{F}/\partial V$, $\partial \mathbf{G}/\partial V$ and $\partial \mathbf{H}/\partial V$ are rather difficult to construct. Appendix A gives explicit forms for $\partial \mathbf{U}/\partial V$ and $\partial \mathbf{F}/\partial V$; the remaining matrices can be constructed via analogy. The inversion of the matrix $\partial \mathbf{U}/\partial V$ has to be done numerically, making the numerical solution of the RMHD equations computationally expensive.

Enforcing eqn. (3.1) at the center of a zone at time $t^n$ in a discrete fashion gives us

$$\Delta_t V = -\frac{\Delta t}{\Delta x} A \left(\Delta_x V\right) - \frac{\Delta t}{\Delta y} B \left(\Delta_y V\right) - \frac{\Delta t}{\Delta z} C \left(\Delta_z V\right) \qquad (3.2)$$

Once $\Delta_t V$ is obtained, the predictor step is done. The time-centered variables that are suited for use within a Riemann solver can now be provided. For example, the primitive variables at the right face and the left face (i.e. the upper x-face and the lower x-face) of the zone being considered can be given as

$$V_R = V + \frac{1}{2}\Delta_x V + \frac{1}{2}\Delta_t V \quad ; \quad V_L = V - \frac{1}{2}\Delta_x V + \frac{1}{2}\Delta_t V \qquad (3.3)$$

This completes our description of the second order accurate predictor.

### III.b) Third Order ADER Scheme

Each cuboidal zone is initially mapped to a unit cube with normalized spatial coordinates $(\chi, \eta, \zeta) \in [-.5, .5]^3$ and a normalized temporal coordinate $\tau \in [0,1]$. This is accomplished with the scaling transformations

$$\chi = x/\Delta x \quad ; \quad \eta = y/\Delta y \quad ; \quad \zeta = z/\Delta z \quad ; \quad \tau = t/\Delta t \qquad (3.4)$$

For structured, hexahedral meshes it is simplest to use tensor products of Legendre polynomials. The spatial polynomials, with domain $[-0.5, 0.5]$, used in this work are given by

$$P_0(\chi) = 1 \; ; \; P_1(\chi) = \chi \; ; \; P_2(\chi) = \chi^2 - \frac{1}{12} \; ; \; P_3(\chi) = \chi^3 - \frac{3}{20}\chi \qquad (3.5)$$

Temporal bases, with domain $[0,1]$, used in this work are given by

$$Q_0(\tau) = 1 \; , \; Q_1(\tau) = \tau \; , \; Q_2(\tau) = \tau^2 \; , \; Q_3(\tau) = \tau^3 \qquad (3.6)$$

In this transformed coordinate system, the governing PDE becomes

$$\frac{\partial V}{\partial \tau} + \frac{\Delta t}{\Delta x} A \frac{\partial V}{\partial \chi} + \frac{\Delta t}{\Delta y} B \frac{\partial V}{\partial \eta} + \frac{\Delta t}{\Delta z} C \frac{\partial V}{\partial \zeta} = 0 \qquad (3.7)$$

The reference element is therefore four-dimensional in space-time and has the domain $[-.5,.5]^3 \times [0,1]$. The third order accurate space-time representation that we seek within the reference element is then given by

$$\begin{aligned}
V(\chi,\eta,\zeta,\tau) &= \hat{v}_1 P_0(\chi) P_0(\eta) P_0(\zeta) Q_0(\tau) \\
&+ \hat{v}_2 P_1(\chi) P_0(\eta) P_0(\zeta) Q_0(\tau) + \hat{v}_3 P_0(\chi) P_1(\eta) P_0(\zeta) Q_0(\tau) + \hat{v}_4 P_0(\chi) P_0(\eta) P_1(\zeta) Q_0(\tau) \\
&+ \hat{v}_5 P_2(\chi) P_0(\eta) P_0(\zeta) Q_0(\tau) + \hat{v}_6 P_0(\chi) P_2(\eta) P_0(\zeta) Q_0(\tau) + \hat{v}_7 P_0(\chi) P_0(\eta) P_2(\zeta) Q_0(\tau) \\
&+ \hat{v}_8 P_1(\chi) P_1(\eta) P_0(\zeta) Q_0(\tau) + \hat{v}_9 P_0(\chi) P_1(\eta) P_1(\zeta) Q_0(\tau) + \hat{v}_{10} P_1(\chi) P_0(\eta) P_1(\zeta) Q_0(\tau) \\
&+ \hat{v}_{11} P_0(\chi) P_0(\eta) P_0(\zeta) Q_1(\tau) + \hat{v}_{12} P_0(\chi) P_0(\eta) P_0(\zeta) Q_2(\tau) \\
&+ \hat{v}_{13} P_1(\chi) P_0(\eta) P_0(\zeta) Q_1(\tau) + \hat{v}_{14} P_0(\chi) P_1(\eta) P_0(\zeta) Q_1(\tau) + \hat{v}_{15} P_0(\chi) P_0(\eta) P_1(\zeta) Q_1(\tau)
\end{aligned}$$

(3.8)

Notice from eqn. (3.8) that the first ten coefficients, from $\hat{v}_1$ to $\hat{v}_{10}$, are provided by a suitable third order accurate spatial reconstruction from the previous section. For example, Balsara *et al.* [15] provide a very efficient WENO strategy for third order accurate reconstruction. The first ten coefficients in eqn. (3.8) do not change in response to the iterative process described here because they do not have temporal dependence; i.e., $Q_0(\tau) = 1$. In the third order accurate ADER iterative process, we seek to evaluate the next five coefficients of eqn. (3.8), from $\hat{v}_{11}$ to $\hat{v}_{15}$, by enforcing consistency with the governing PDE in eqn. (3.7). In other words, notice from our definitions in eqn. (3.6) that those five coefficients carry time-dependence because $Q_1(\tau) = \tau$ and $Q_2(\tau) = \tau^2$.

We can always start the iterative ADER process by initializing $\hat{v}_{11} = \hat{v}_{12} = \hat{v}_{13} = \hat{v}_{14} = \hat{v}_{15} = 0$. The important task is to obtain $\hat{v}_{11}$ to $\hat{v}_{15}$. This is accomplished by picking five judiciously designed nodal points in the reference space-time element and enforcing consistency with the governing PDE at those nodal points. At third order, we pick the six nodal points in space-time given by

$$n_1 \equiv [0,0,0,0], \ n_2 \equiv [1/2,0,0,0], \ n_3 \equiv [0,1/2,0,0], \ n_4 \equiv [0,0,1/2,0],$$
$$n_5 \equiv [0,0,0,1/2], \ n_6 \equiv [0,0,0,1]$$

(3.9)

At any stage in the iterative process, eqn. (3.8) is capable of giving us V, $\partial V/\partial \chi$, $\partial V/\partial \eta$ and $\partial V/\partial \zeta$ at any space-time point inside the reference element. Of course, we choose to evaluate those terms at the nodal points from eqn. (3.9). Thus at the $i^{th}$ nodal point, with $i = 1,...,6$, we evaluate $V_{ni}$, $(\partial V/\partial \chi)_{ni}$, $(\partial V/\partial \eta)_{ni}$ and $(\partial V/\partial \zeta)_{ni}$. At each nodal point, the vector of primitives "$V_{ni}$" can then be used to build the matrices "A", "B" and "C" in eqn. (3.7). Consequently, at each of the nodal points, we can obtain $(\partial V/\partial \tau)_{ni}$ from the governing eqn. (3.7). But note that this $(\partial V/\partial \tau)_{ni}$ which is obtained from the governing equation can also be equated to an analogous equation that is obtained by differentiating eqn. (3.8) with respect to "$\tau$" at the same nodal point. This is how consistency with the governing PDE is enforced. Enforcing this consistency enables us to fix up the coefficients $\hat{v}_{11}$ to $\hat{v}_{15}$. This task is explicitly carried out in the next five equations.

$$\hat{v}_{11} = (\partial V/\partial \tau)_{n1} \tag{3.10}$$

$$\hat{v}_{13} = 2\left((\partial V/\partial \tau)_{n2} - (\partial V/\partial \tau)_{n1}\right) \tag{3.11}$$

$$\hat{v}_{14} = 2\left((\partial V/\partial \tau)_{n3} - (\partial V/\partial \tau)_{n1}\right) \tag{3.12}$$

$$\hat{v}_{15} = 2\left((\partial V/\partial \tau)_{n4} - (\partial V/\partial \tau)_{n1}\right) \tag{3.14}$$

$$\hat{v}_{12} = \frac{1}{2}\left[4(\partial V/\partial \tau)_{n5} - (\partial V/\partial \tau)_{n6} - 3(\partial V/\partial \tau)_{n1}\right] \tag{3.15}$$

While this completes our formal description of the third order ADER predictor step, the next paragraph shows how the ADER iteration is implemented in practice.

From the temporal arrangement of the nodal points in eqn. (3.9) please observe that for nodes $n_1$ to $n_4$ we only need to evaluate $(\partial V/\partial \tau)_{n1}$ to $(\partial V/\partial \tau)_{n4}$ once. In other words, the governing equation only needs to be invoked and evaluated at those nodal points once and only once. From eqns. (3.10) to (3.14) we see that this invocation of the governing PDE is sufficient to fix up $\hat{v}_{11}$, $\hat{v}_{13}$, $\hat{v}_{14}$ and $\hat{v}_{15}$. Only nodes $n_5$ and $n_6$ need to participate in the iterative process and we see from eqn. (3.15) that they enable us to progressively improve the value of $\hat{v}_{12}$. The ADER iteration is a form of Picard iteration, in other words, a fixed point iteration. There is a substantial body of experience indicating that to obtain the five unknown coefficients in eqn. (3.8) with third order of accuracy, only two iterations are needed (see Dumbser *et al.* [33]). Thus the convergence is very fast.

### III.c) Fourth Order ADER Scheme

Eqns. (3.4) to (3.7) continue unchanged in the fourth order case. The analogue of eqn. (3.8) now becomes

$$
\begin{aligned}
V(\chi,\eta,\zeta,\tau) &= \hat{v}_1 P_0(\chi) P_0(\eta) P_0(\zeta) Q_0(\tau) \\
&+ \hat{v}_2 P_1(\chi) P_0(\eta) P_0(\zeta) Q_0(\tau) + \hat{v}_3 P_0(\chi) P_1(\eta) P_0(\zeta) Q_0(\tau) + \hat{v}_4 P_0(\chi) P_0(\eta) P_1(\zeta) Q_0(\tau) \\
&+ \hat{v}_5 P_2(\chi) P_0(\eta) P_0(\zeta) Q_0(\tau) + \hat{v}_6 P_0(\chi) P_2(\eta) P_0(\zeta) Q_0(\tau) + \hat{v}_7 P_0(\chi) P_0(\eta) P_2(\zeta) Q_0(\tau) \\
&+ \hat{v}_8 P_1(\chi) P_1(\eta) P_0(\zeta) Q_0(\tau) + \hat{v}_9 P_0(\chi) P_1(\eta) P_1(\zeta) Q_0(\tau) + \hat{v}_{10} P_1(\chi) P_0(\eta) P_1(\zeta) Q_0(\tau) \\
&+ \hat{v}_{11} P_3(\chi) P_0(\eta) P_0(\zeta) Q_0(\tau) + \hat{v}_{12} P_0(\chi) P_3(\eta) P_0(\zeta) Q_0(\tau) + \hat{v}_{13} P_0(\chi) P_0(\eta) P_3(\zeta) Q_0(\tau) \\
&+ \hat{v}_{14} P_2(\chi) P_1(\eta) P_0(\zeta) Q_0(\tau) + \hat{v}_{15} P_2(\chi) P_0(\eta) P_1(\zeta) Q_0(\tau) \\
&+ \hat{v}_{16} P_1(\chi) P_2(\eta) P_0(\zeta) Q_0(\tau) + \hat{v}_{17} P_0(\chi) P_2(\eta) P_1(\zeta) Q_0(\tau) \\
&+ \hat{v}_{18} P_1(\chi) P_0(\eta) P_2(\zeta) Q_0(\tau) + \hat{v}_{19} P_0(\chi) P_1(\eta) P_2(\zeta) Q_0(\tau) \\
&+ \hat{v}_{20} P_1(\chi) P_1(\eta) P_1(\zeta) Q_0(\tau) \\
&+ \hat{v}_{21} P_0(\chi) P_0(\eta) P_0(\zeta) Q_1(\tau) + \hat{v}_{22} P_0(\chi) P_0(\eta) P_0(\zeta) Q_2(\tau) + \hat{v}_{23} P_0(\chi) P_0(\eta) P_0(\zeta) Q_3(\tau) \\
&+ \hat{v}_{24} P_1(\chi) P_0(\eta) P_0(\zeta) Q_1(\tau) + \hat{v}_{25} P_0(\chi) P_1(\eta) P_0(\zeta) Q_1(\tau) + \hat{v}_{26} P_0(\chi) P_0(\eta) P_1(\zeta) Q_1(\tau) \\
&+ \hat{v}_{27} P_1(\chi) P_0(\eta) P_0(\zeta) Q_2(\tau) + \hat{v}_{28} P_0(\chi) P_1(\eta) P_0(\zeta) Q_2(\tau) + \hat{v}_{29} P_0(\chi) P_0(\eta) P_1(\zeta) Q_2(\tau) \\
&+ \hat{v}_{30} P_2(\chi) P_0(\eta) P_0(\zeta) Q_1(\tau) + \hat{v}_{31} P_0(\chi) P_2(\eta) P_0(\zeta) Q_1(\tau) + \hat{v}_{32} P_0(\chi) P_0(\eta) P_2(\zeta) Q_1(\tau) \\
&+ \hat{v}_{33} P_1(\chi) P_1(\eta) P_0(\zeta) Q_1(\tau) + \hat{v}_{34} P_0(\chi) P_1(\eta) P_1(\zeta) Q_1(\tau) + \hat{v}_{35} P_1(\chi) P_0(\eta) P_1(\zeta) Q_1(\tau)
\end{aligned}
$$

(3.16)

Notice from eqn. (3.16) that the first twenty coefficients, from $\hat{v}_1$ to $\hat{v}_{20}$, are provided by a suitable fourth order accurate spatial reconstruction from the previous section. For example, Balsara *et al.* [15] provide a very efficient WENO strategy for fourth order accurate reconstruction. The first twenty coefficients in eqn. (3.16) do not change in response to the iterative process described here. In the fourth order accurate ADER iterative process, we seek to evaluate the next five coefficients of eqn. (3.16), from $\hat{v}_{21}$ to $\hat{v}_{35}$, by enforcing consistency with the governing PDE in eqn. (3.7).

We can start the fourth order ADER iteration by initializing $\hat{v}_{21}$ to $\hat{v}_{35}$ to zero. The important task is to obtain $\hat{v}_{21}$ to $\hat{v}_{35}$. This is accomplished by picking fifteen judiciously designed nodal points in the reference space-time element and enforcing consistency with the governing PDE at those nodal points. At fourth order, we pick the fifteen nodal points in space-time given by

$$n_1 \equiv [0,0,0,0],\ n_2 \equiv [1/2,0,0,0],\ n_3 \equiv [-1/2,0,0,0],\ n_4 \equiv [0,1/2,0,0],\ n_5 \equiv [0,-1/2,0,0],$$
$$n_6 \equiv [0,0,1/2,0],\ n_7 \equiv [0,0,-1/2,0],\ n_8 \equiv [0,0,0,1/2],\ n_9 \equiv [0,0,0,1],\ n_{10} \equiv [1/2,1/2,0,1],$$
$$n_{11} \equiv [-1/2,-1/2,0,1],\ n_{12} \equiv [1/2,0,1/2,1],\ n_{13} \equiv [-1/2,0,-1/2,1],\ n_{14} \equiv [0,1/2,1/2,1],$$
$$n_{15} \equiv [0,-1/2,-1/2,1]$$

(3.17)

Because our fourth order strategy mimics the one described in detail for the third order case above, we do not repeat details. Thus at the $i^{th}$ nodal point, with $i = 1,...,15$, we evaluate $V_{ni}$, $(\partial V/\partial \chi)_{ni}$, $(\partial V/\partial \eta)_{ni}$ and $(\partial V/\partial \zeta)_{ni}$. At each nodal point, the vector of primitives "$V_{ni}$" can then be used to build the matrices "A", "B" and "C" in eqn. (3.7). Consequently, at each of the nodal points, we can obtain $(\partial V/\partial \tau)_{ni}$ from the governing eqn. (3.7). Seven out of the fifteen coefficients that we seek can be evaluated by using the nodes which have $\tau = 0$. From those seven nodes, we can evaluate the seven coefficients $\hat{v}_{21}$, $\hat{v}_{24}$, $\hat{v}_{25}$, $\hat{v}_{26}$, $\hat{v}_{30}$, $\hat{v}_{31}$ and $\hat{v}_{32}$ once and for all by using the seven equations below.

$$\hat{v}_{21} = \left((\partial V/\partial \tau)_{n2} + (\partial V/\partial \tau)_{n3} + (\partial V/\partial \tau)_{n4} + (\partial V/\partial \tau)_{n5} + (\partial V/\partial \tau)_{n6} + (\partial V/\partial \tau)_{n7}\right)/6 \quad (3.18)$$

$$\hat{v}_{24} = (\partial V/\partial \tau)_{n2} - (\partial V/\partial \tau)_{n3} \quad (3.19)$$

$$\hat{v}_{25} = (\partial V/\partial \tau)_{n4} - (\partial V/\partial \tau)_{n5} \quad (3.20)$$

$$\hat{v}_{26} = (\partial V/\partial \tau)_{n6} - (\partial V/\partial \tau)_{n7} \quad (3.21)$$

$$\hat{v}_{30} = 2\left(\left(\partial V/\partial \tau\right)_{n2} + \left(\partial V/\partial \tau\right)_{n3} - 2\left(\partial V/\partial \tau\right)_{n1}\right) \tag{3.22}$$

$$\hat{v}_{31} = 2\left(\left(\partial V/\partial \tau\right)_{n4} + \left(\partial V/\partial \tau\right)_{n5} - 2\left(\partial V/\partial \tau\right)_{n1}\right) \tag{3.23}$$

$$\hat{v}_{32} = 2\left(\left(\partial V/\partial \tau\right)_{n6} + \left(\partial V/\partial \tau\right)_{n7} - 2\left(\partial V/\partial \tau\right)_{n1}\right) \tag{3.24}$$

The other eight coefficients in eqn. (3.16) are obtained via iteration. Each ADER iteration improves the quality of the space-time solution in eqn. (3.16). The equations for $\hat{v}_{22}$, $\hat{v}_{23}$, $\hat{v}_{27}$, $\hat{v}_{28}$, $\hat{v}_{29}$, $\hat{v}_{33}$, $\hat{v}_{34}$ and $\hat{v}_{35}$ are given by

$$\hat{v}_{22} = 2\left(\partial V/\partial \tau\right)_{n8} - \frac{1}{2}\left(\partial V/\partial \tau\right)_{n9} - \frac{3}{2}V_{n21} + \frac{1}{8}\left(V_{n30} + V_{n31} + V_{n32}\right) \tag{3.25}$$

$$\hat{v}_{23} = \frac{2}{3}\left(-2\left(\partial V/\partial \tau\right)_{n8} + \left(\partial V/\partial \tau\right)_{n9} + V_{n21}\right) - \frac{1}{18}\left(V_{n30} + V_{n31} + V_{n32}\right) \tag{3.26}$$

$$\hat{v}_{27} = \left(\left(\partial V/\partial \tau\right)_{n10} - \left(\partial V/\partial \tau\right)_{n11} + \left(\partial V/\partial \tau\right)_{n12} - \left(\partial V/\partial \tau\right)_{n13} - \left(\partial V/\partial \tau\right)_{n14} + \left(\partial V/\partial \tau\right)_{n15}\right)/4 - V_{n24}/2 \tag{3.27}$$

$$\hat{v}_{28} = \left(\left(\partial V/\partial \tau\right)_{n10} - \left(\partial V/\partial \tau\right)_{n11} - \left(\partial V/\partial \tau\right)_{n12} + \left(\partial V/\partial \tau\right)_{n13} + \left(\partial V/\partial \tau\right)_{n14} - \left(\partial V/\partial \tau\right)_{n15}\right)/4 - V_{n25}/2 \tag{3.28}$$

$$\hat{v}_{29} = \left(-\left(\partial V/\partial \tau\right)_{n10} + \left(\partial V/\partial \tau\right)_{n11} + \left(\partial V/\partial \tau\right)_{n12} - \left(\partial V/\partial \tau\right)_{n13} + \left(\partial V/\partial \tau\right)_{n14} - \left(\partial V/\partial \tau\right)_{n15}\right)/4 - V_{n26}/2 \tag{3.29}$$

$$\hat{v}_{33} = 2\left(\left(\partial V/\partial \tau\right)_{n10} + \left(\partial V/\partial \tau\right)_{n11} - 2\left(\partial V/\partial \tau\right)_{n9}\right) - \left(V_{n30} + V_{n31}\right) \tag{3.30}$$

$$\hat{v}_{34} = 2\left(\left(\partial V/\partial \tau\right)_{n14} + \left(\partial V/\partial \tau\right)_{n15} - 2\left(\partial V/\partial \tau\right)_{n9}\right) - \left(V_{n31} + V_{n32}\right) \tag{3.31}$$

$$\hat{v}_{35} = 2\left(\left(\partial V/\partial \tau\right)_{n12} + \left(\partial V/\partial \tau\right)_{n13} - 2\left(\partial V/\partial \tau\right)_{n9}\right) - \left(V_{n30} + V_{n32}\right) \tag{3.32}$$

This completes our formal description of the fourth order ADER predictor step. As before, the process converges very fast and we only need to carry out three iterations of eqns. (3.25) to (3.31) to realize fourth order of accuracy.

The careful reader should also be cautioned that recently Zanotti and Dumbser [71] have advocated the use of ADER-based time evolution of the primitive variables given by $V = \left(\rho, v_x, v_y, v_z, P_g, B_x, B_y, B_z\right)^T$. This choice of primitive variables has no special advantage with respect to retaining subluminal velocities in RMHD. This point was also made in our talks at the HONOM meeting in Trento on March 2015 and the Astronum meeting in Avignon on June 2015; and it seems that the point was not appreciated. Furthermore, please note that all aspects of the ADER predictor step are explicitly provided here in a format that is very efficient for implementation.

**IV) The Corrector Step: Formulation of the Multidimensional Riemann Solver for RMHD**

We assume that the RMHD equations are being solved with mass, momentum and energy being zone-centered and the magnetic variables collocated on a staggered Yee-type mesh. Consequently, the magnetic field components are collocated at the face centers of the mesh and the electric field components are collocated at the edge centers of the mesh. The reconstruction strategy from Section II has produced the spatial variation of primitive variables and the predictor step from Section III has provided an "in-the-small" evolution of the PDE system within a zone. The Riemann solvers now have to supply the corrector step. This couples all the zones to their neighbors during the timestep. In sub-section IV.a we briefly describe modern advances in one-dimensional Riemann solvers for RMHD. In sub-section IV.b we describe the multidimensional Riemann solver for RMHD.

**IV.a) Brief Description of Recent Advances in One-dimensional Riemann solvers for RMHD**

The RMHD system is so computationally expensive that only the simplest one-dimensional Riemann solvers are used in practice. These one-dimensional Riemann solvers are needed as building blocks for the multidimensional Riemann solver. They are also needed in the evaluation of the facially-averaged fluxes at zone boundaries for conserved variables. Because RMHD is computationally expensive, HLL and HLLC Riemann solvers for RMHD are the only ones that have seen maximal use. These HLLC Riemann solvers for RMHD (Mignone & Bodo [56], Kim & Balsara [49]) follow the pattern set for analogous HLLC/HLLD Riemann solvers that have been designed for MHD (Gurski [46], Li [52], Miyoshi & Kusano [58]). Toro, Spruce and Speares [68] and Batten *et al*. [25] have shown that it is essential to maintain constancy of pressure and longitudinal velocity in the resolved states of the HLLC Riemann solver. Appendix

B of this paper presents a modified version of the HLLC Riemann solver for RMHD which incorporates this essential insight from Toro, Spruce and Speares [68].

Recently, Dumbser and Balsara [36], Balsara *et al.* [23] have introduced a one-dimensional HLLEM Riemann solver which also shares all the same positivity properties as the one-dimensional HLL/HLLC Riemann solvers. Our new HLLEM Riemann solver works seamlessly for RMHD. With the HLLC Riemann solver, it shares the virtue that it can capture isolated, stationary contact discontinuities. However, it goes beyond the HLLC Riemann solver in that it can also capture isolated, stationary Alfvenic discontinuities. The HLLD Riemann solver for RMHD (Mignone, Ugliano and Bodo [57]) can also capture isolated, stationary Alfven waves; however, is very expensive because of its iterative nature. It is also very brittle in situations where the iterations fail to converge. The new HLLEM Riemann solver has the advantage of not requiring any iterations. Furthermore, it just requires an evaluation of only a subset of the eigenvectors that are needed. For the one-dimensional as well as the multidimensional Riemann solvers used here, we have only used the entropy wave eigenvector along with the Alfven wave eigenvectors because they are the only linearly degenerate waves that need to be represented crisply on the mesh. Furthermore, realize that the eigenvectors for the entropy wave and the Alfven waves are easy to evaluate for RMHD (Balsara [7], Anton *et al.* [4]).

**IV.b) Description of the Multidimensional Riemann solver for RMHD**

The motivation for a multidimensional Riemann solver for RMHD is easily seen from eqn. (2.1). Notice that the sixth component of the y-flux is the opposite of the seventh component of the x-flux and is equal to the z-component of the electric field. The upwinded z-component of the electric field has, therefore, to take suitably upwinded contributions from the x-directional and y-directional wave motions. One dimensional Riemann solvers cannot arbitrate the multidimensionally upwinded direction. Solving the multidimensional Riemann problem is indeed the correct way to obtain the multidimensionally upwinded electric fields at the edges of the mesh. For MHD and RMHD, the upwinding only needs to be two dimensional because the flux in the third dimension does not contribute to the electric field. For example, the z-flux in eqn. (2.1) does not contribute to the z-component of the electric field.

Because most applications of RMHD are likely to be on a structured mesh, we show how a multidimensional Riemann solver operates in two dimensions on a structured mesh. Realize that we only get a multidimensional view of wave propagation at the vertex of a mesh. Therefore, Fig. 1a depicts four states $\mathbf{U}_{RU}, \mathbf{U}_{LU}, \mathbf{U}_{RD}$ and $\mathbf{U}_{LD}$ that come together at the vertex of a mesh, shown by the gray dot. These are the input states for a multidimensional Riemann solver. As a helpful mnemonic, the subscripts "R" and "L" stand for right and left respectively; the subscripts "U" and "D" stand for up and down respectively. Between the states $\mathbf{U}_{RU}$ and $\mathbf{U}_{LU}$, a one-dimensional Riemann problem propagates in the x-direction. Similarly, between the states $\mathbf{U}_{RD}$ and $\mathbf{U}_{LD}$, a one-dimensional Riemann problem propagates in the x-direction. Between the states $\mathbf{U}_{RU}$ and $\mathbf{U}_{RD}$, a one-dimensional Riemann problem propagates in the y-direction. Likewise, between the states $\mathbf{U}_{LU}$ and $\mathbf{U}_{LD}$, a one-dimensional Riemann problem propagates in the y-direction. Fig. 1b is a schematic diagram where the thick black lines show the propagation of the one dimensional waves from the one-dimensional HLLC Riemann solvers. As a result, between the two extremal waves of each one-dimensional Riemann problem we have one intermediate wave and two intermediate states. All these one-dimensional Riemann problems interact with each other at the vertex. The interaction of multiple one-dimensional Riemann problems gives rise to a strongly interacting state. This strongly-interacting state (denoted S.I. State in Fig. 1b) is shown by the rectangle with thick black dashed lines in Fig. 1b. Understanding the interior structure of the strongly interacting state is crucial for getting a good solution of the subsonic, multidimensional Riemann problem. The black dashed lines in Fig. 1b are referred to as the multidimensional wave model so that we can say that the multidimensional wave model bounds the strongly-interacting state. As shown in Balsara [14], entropy enforcement considerations enable us to claim that the strongly interacting state is bounded by extremal speeds that are provided by the one-dimensional Riemann problems. Eqn. (4) and (7) of Balsara [14] enables us to find those bounding speeds so that we can say that the strongly-interacting state is bounded (in the space of speeds) by $[S_L, S_R] \times [S_D, S_U]$. These speeds are also shown in Fig. 1b.

The one-dimensional Riemann problems always evolve self-similarly and it was shown in Schulz-Rinne *et al*. [63] that the strongly-interacting state also evolves self-similarly. While the self-similarity of the strongly-interacting state has been implicitly built into the formulation of the multidimensional Riemann problem, it was used with even greater effect in Balsara [17] who showed that similarity variables could be used to obtain the sub-structure of the strongly-interacting state. Balsara [17] provides an extensive discussion on how to obtain the sub-structure of the multidimensional Riemann problem. In this paper we provide just the minimal ideas that are needed for implementation of a MuSIC Riemann solver on a Cartesian mesh.

We assume that we are solving at a z-edge of the mesh and that the four states shown in Fig. 1a are the input RMHD states to our MuSIC Riemann solver. The similarity variables are given by $\tilde{\xi} \equiv x/t$ and $\tilde{\psi} \equiv y/t$ so that the extent of the strongly interacting state in Fig. 1b is given by $(\tilde{\xi},\tilde{\psi}) \in [S_L, S_R] \times [S_D, S_U]$. The analysis from Balsara [17] and Balsara *et al.* [23] shows that it is physically most meaningful to express the variation in the strongly-interacting state in shifted similarity variables. We make a coordinate transformation in the similarity variables with

$$\xi_c \equiv (S_R + S_L)/2 \quad ; \quad \Delta\xi \equiv (S_R - S_L) \quad ; \quad \psi_c \equiv (S_U + S_D)/2 \quad ; \quad \Delta\psi \equiv (S_U - S_D)$$
$$\xi \equiv \left(\frac{\tilde{\xi} - \xi_c}{\Delta\xi}\right) \quad ; \quad \psi \equiv \left(\frac{\tilde{\psi} - \psi_c}{\Delta\psi}\right) \tag{4.1}$$

Observe that $\xi$ and $\psi$ are still self-similar variables with the main difference that they now range over $[-1/2, 1/2] \times [-1/2, 1/2]$. The centroid of the rectangle that bounds the strongly-interacting state is shown by the black dot in Fig. 1b and the shifted similarity variables are also shown. From here on, we refer to the shifted similarity variables as just the similarity variables. Notice that the states associated with the four one-dimensional Riemann solvers in Fig. 1b provide the boundary values of our strongly interacting state. To take an example, the coordinate $(1/2, \psi)$ allows us to pick out the states associated with the one-dimensional Riemann problem that is formed between $\mathbf{U}_{RU}$ and $\mathbf{U}_{RD}$. As $\psi$ ranges from -1/2 to 1/2, with $\xi$ set to $1/2$, we can pick out all the states $\mathbf{U}_{RU}, \mathbf{U}_R^{*-}, \mathbf{U}_R^{*+}$, and $\mathbf{U}_{RD}$ in Fig. 1b. The fluxes associated with those states can also be picked out. The intermediate states in the other three Riemann problems in Fig. 1b are not shown because we want to keep the figure reasonably uncluttered.

For numerical use, it is fully adequate to retain linear sub-structure in the strongly-interacting state. Thus we expand the strongly interacting state about the centroid of our wave model. We do this using our similarity variables as

$$\tilde{\mathbf{U}}(\xi,\psi) = \bar{\mathbf{U}} + \mathbf{U}_\xi \xi + \mathbf{U}_\psi \psi \tag{4.2}$$

Here $\bar{\mathbf{U}}$ is the mean state in the multidimensional Riemann solver; while $\mathbf{U}_\xi$ and $\mathbf{U}_\psi$ carry the multidimensional sub-structure in the multidimensional Riemann solver. The inclusion of sub-structure helps in reducing dissipation. A similar expansion can be made for the x-flux as

$$\tilde{\mathbf{F}}(\xi,\psi) = \bar{\mathbf{F}} + \bar{\mathbf{A}}\left(\mathbf{U}_\xi \xi + \mathbf{U}_\psi \psi\right) \quad \text{with} \quad \bar{\mathbf{A}} = \frac{\partial \mathbf{F}(\bar{\mathbf{U}})}{\partial \bar{\mathbf{U}}} \tag{4.3}$$

and for the y-flux as

$$\tilde{\mathbf{G}}(\xi,\psi) = \bar{\mathbf{G}} + \bar{\mathbf{B}}\left(\mathbf{U}_\xi \xi + \mathbf{U}_\psi \psi\right) \quad \text{with} \quad \bar{\mathbf{B}} = \frac{\partial \mathbf{G}(\bar{\mathbf{U}})}{\partial \bar{\mathbf{U}}} \tag{4.4}$$

In the next paragraph we provide sufficient detail that would enable the implementer to obtain the terms in eqns. (4.2), (4.3) and (4.4).

It can be shown, see Balsara [17], that taking the zeroth moment of the conservation law over the entire multidimensional wave model gives

$$\bar{\mathbf{U}} = -\frac{1}{2}\left[\begin{array}{l}\frac{1}{\Delta\xi}\int_{-1/2}^{1/2}\left(\mathbf{F}(1/2,\psi) - S_R \mathbf{U}(1/2,\psi)\right)d\psi - \frac{1}{\Delta\xi}\int_{-1/2}^{1/2}\left(\mathbf{F}(-1/2,\psi) - S_L \mathbf{U}(-1/2,\psi)\right)d\psi \\ +\frac{1}{\Delta\psi}\int_{-1/2}^{1/2}\left(\mathbf{G}(\xi,1/2) - S_U \mathbf{U}(\xi,1/2)\right)d\xi - \frac{1}{\Delta\psi}\int_{-1/2}^{1/2}\left(\mathbf{G}(\xi,-1/2) - S_D \mathbf{U}(\xi,-1/2)\right)d\xi\end{array}\right] \tag{4.5}$$

Notice that the right hand side of the above equation is not a matrix. The states and fluxes on the right hand side of eqn. (4.5) are provided by the one dimensional Riemann solvers that reside on the outer boundary of the multidimensional wave model. The above equation shows that the mean conserved state, $\bar{\mathbf{U}}$, in the multidimensional wave model is just the sum of the Lagrangian fluxes that flow through the boundaries of the wave model. This makes physical sense because the wave model is a self-similarly expanding boundary and the Lagrangian fluxes account for this expansion. Here $\bar{\mathbf{U}}$ is the mean value of the strongly-interacting state and it is the first thing that is evaluated in a multidimensional Riemann solver. Balsara [17] shows that taking the $\xi$-moment of the conservation law over the entire multidimensional wave model gives

$$\bar{\mathbf{F}} = \xi_c \bar{\mathbf{U}} + \Delta\xi \left[ \begin{array}{l} \dfrac{1}{2\Delta\xi} \int_{-1/2}^{1/2} \left( \mathbf{F}(1/2,\psi) - S_R \mathbf{U}(1/2,\psi) \right) d\psi + \dfrac{1}{2\Delta\xi} \int_{-1/2}^{1/2} \left( \mathbf{F}(-1/2,\psi) - S_L \mathbf{U}(-1/2,\psi) \right) d\psi \\ + \dfrac{1}{\Delta\psi} \int_{-1/2}^{1/2} \xi \left( \mathbf{G}(\xi,1/2) - S_U \mathbf{U}(\xi,1/2) \right) d\xi - \dfrac{1}{\Delta\psi} \int_{-1/2}^{1/2} \xi \left( \mathbf{G}(\xi,-1/2) - S_D \mathbf{U}(\xi,-1/2) \right) d\xi \end{array} \right]$$
$$+ \dfrac{\Delta\xi}{4} \mathbf{U}_\xi$$

(4.6)

Similarly, taking the $\psi$-moment of the conservation law over the entire multidimensional wave model gives

$$\bar{\mathbf{G}} = \psi_c \bar{\mathbf{U}} + \Delta\psi \left[ \begin{array}{l} \dfrac{1}{\Delta\xi} \int_{-1/2}^{1/2} \psi \left( \mathbf{F}(1/2,\psi) - S_R \mathbf{U}(1/2,\psi) \right) d\psi - \dfrac{1}{\Delta\xi} \int_{-1/2}^{1/2} \psi \left( \mathbf{F}(-1/2,\psi) - S_L \mathbf{U}(-1/2,\psi) \right) d\psi \\ + \dfrac{1}{2\Delta\psi} \int_{-1/2}^{1/2} \left( \mathbf{G}(\xi,1/2) - S_U \mathbf{U}(\xi,1/2) \right) d\xi + \dfrac{1}{2\Delta\psi} \int_{-1/2}^{1/2} \left( \mathbf{G}(\xi,-1/2) - S_D \mathbf{U}(\xi,-1/2) \right) d\xi \end{array} \right]$$
$$+ \dfrac{\Delta\psi}{4} \mathbf{U}_\psi$$

(4.7)

This is the simplest equation set needed for making an adequately nice implementation of a multidimensional Riemann solver that retains sub-structure. We describe the process of retaining sub-structure in the next paragraph.

Realize that the state $\bar{\mathbf{U}}$ will always be available via eqn. (4.5) and we start by obtaining it first. If we set $\mathbf{U}_\xi = \mathbf{U}_\psi = 0$, i.e. we suppress substructure, then eqns. (4.6) and (4.7) immediately give us the multidimensionally upwinded numerical fluxes $\bar{\mathbf{F}}$ and $\bar{\mathbf{G}}$. This yields the most stable multidimensional Riemann solver and if one is satisfied with a slightly greater dissipation, one can just set $\mathbf{U}_\xi = \mathbf{U}_\psi = 0$ and obtain the state $\bar{\mathbf{U}}$ as well as the fluxes $\bar{\mathbf{F}}$ and $\bar{\mathbf{G}}$ from the multidimensional Riemann solver. Such a multidimensional Riemann solver would be the analogue of the one-dimensional HLL Riemann solver. Like the one-dimensional Riemann solver, it lacks substructure and therefore has a somewhat higher level of dissipation. If this Riemann solver is used, it will increase dissipation everywhere, not just at shocks where the added dissipation is needed. In Balsara *et al*. [23] a great deal of analysis has been carried out which shows us how to reduce dissipation by introducing sub-structure into the multidimensional Riemann solver. While we do not repeat the mathematical details and logic here, we provide sufficient information to make this section self-contained and facilitate implementation.

To introduce sub-structure, we have to first introduce undivided differences which can be evaluated by using the solutions to the one-dimensional Riemann problems at the boundaries of the wave model. In other words, we first evaluate

$$\left(\Delta_\xi \mathbf{U}\right) = \int_{-1/2}^{1/2} \mathbf{U}(1/2,\psi)\,d\psi - \int_{-1/2}^{1/2} \mathbf{U}(-1/2,\psi)\,d\psi \tag{4.8}$$

and

$$\left(\Delta_\psi \mathbf{U}\right) = \int_{-1/2}^{1/2} \mathbf{U}(\xi,1/2)\,d\xi - \int_{-1/2}^{1/2} \mathbf{U}(\xi,-1/2)\,d\xi \tag{4.9}$$

It is useful to have a criterion that tells us when the introduction of a linear profile is acceptable and when it isn't. In Balsara [17] such a criterion was built based on the multidimensional limiter of Barth & Frederickson [24]. Appendix C of Balsara [17] describes the criterion in detail.

The undivided differences $\left(\Delta_\xi \mathbf{U}\right)$ and $\left(\Delta_\psi \mathbf{U}\right)$ from the previous paragraph can then be related to $\mathbf{U}_\xi$ and $\mathbf{U}_\psi$ respectively by appropriate projections that can be made with the left and right eigenvectors. To that end, we identify the interior waves in both directions for the state $\bar{\mathbf{U}}$. Let $\left\{\lambda_i^x : i \in I_{\text{int}}\right\}$, $\left\{r_i^x : i \in I_{\text{int}}\right\}$ and $\left\{l_i^x : i \in I_{\text{int}}\right\}$ be the eigenvalues and right- and left-eigenvectors in the x-direction associated with the state $\bar{\mathbf{U}}$. Likewise, let $\left\{\lambda_i^y : i \in I_{\text{int}}\right\}$, $\left\{r_i^y : i \in I_{\text{int}}\right\}$ and $\left\{l_i^y : i \in I_{\text{int}}\right\}$ be the eigenvalues and right- and left-eigenvectors in the y-direction associated with the state $\bar{\mathbf{U}}$. We assume that the eigenstates are so ordered that the same set $I_{\text{int}}$ labels the intermediate waves in either direction; this is usually possible for most hyperbolic systems. (For the RMHD work reported here, the set $I_{\text{int}}$ labels the entropy wave and the two Alfven waves.) We can now relate $\mathbf{U}_\xi$ to $\left(\Delta_\xi \mathbf{U}\right)$ as follows

$$\mathbf{U}_\xi = \sum_{i \in I_{int}} \left[ 2\delta_i^x \ l_i^x \bullet \left( \Delta_\xi \mathbf{U} \right) \right] r_i^x \tag{4.10}$$

We can also relate $\mathbf{U}_\psi$ to $\left( \Delta_\psi \mathbf{U} \right)$ as

$$\mathbf{U}_\psi = \sum_{i \in I_{int}} \left[ 2\delta_i^y \ l_i^y \bullet \left( \Delta_\psi \mathbf{U} \right) \right] r_i^y \tag{4.11}$$

To fully specify eqns. (4.10) and (4.11), we need to specify $\delta_i^x$ and $\delta_i^y$ for all the waves of interest. This is achieved when

$$\delta_i^x = \begin{cases} \phi_i^x & \text{when } \left[ (S_R - S_L)^2 / 2 - \lambda_i^x (S_R + S_L) \right] \leq 0 \\ \min \left( \dfrac{S_R \lambda_i^{x-} + S_L \lambda_i^{x+} - S_R S_L}{\left[ (S_R - S_L)^2 / 2 - \lambda_i^x (S_R + S_L) \right]}, \phi_i^x \right) & \text{otherwise} \end{cases} \tag{4.12}$$

where $\phi_i^x = -\dfrac{2 S_R S_L}{(S_R - S_L)^2}$

and

$$\delta_i^y = \begin{cases} \phi_i^y & \text{when } \left[ (S_U - S_D)^2 / 2 - \lambda_i^y (S_U + S_D) \right] \leq 0 \\ \min \left( \dfrac{S_U \lambda_i^{y-} + S_D \lambda_i^{y+} - S_U S_D}{\left[ (S_U - S_D)^2 / 2 - \lambda_i^y (S_U + S_D) \right]}, \phi_i^y \right) & \text{otherwise} \end{cases} \tag{4.13}$$

where $\phi_i^y = -\dfrac{2 S_U S_D}{(S_U - S_D)^2}$

Thus the strategy for introducing sub-structure is to first evaluate the undivided differences $\left( \Delta_\xi \mathbf{U} \right)$ and $\left( \Delta_\psi \mathbf{U} \right)$ using eqns. (4.8) and (4.9). Then limit the profiles as needed using Appendix C of Balsara [17]. Then specify $\delta_i^x$ and $\delta_i^y$ using eqns. (4.12) and (4.13). We are now ready to evaluate $\mathbf{U}_\xi$ and $\mathbf{U}_\psi$ using eqns. (4.10) and (4.11).

Once $\mathbf{U}_\xi$ and $\mathbf{U}_\psi$ have been evaluated using the narrative in the previous paragraph, eqns. (4.6) and (4.7) give us $\overline{\mathbf{F}}$ and $\overline{\mathbf{G}}$ with the inclusion of sub-structure. The last step consists of evaluating the numerical fluxes with sub-structure that are given at the vertices of the mesh. This is given explicitly in the next two formulae

$$\mathbf{F}_{numerical} = \overline{\mathbf{F}} - \left(\frac{\xi_c}{\Delta \xi}\right) \overline{\mathbf{A}}\ \mathbf{U}_\xi \qquad (4.14)$$

and

$$\mathbf{G}_{numerical} = \overline{\mathbf{G}} - \left(\frac{\psi_c}{\Delta \psi}\right) \overline{\mathbf{B}}\ \mathbf{U}_\psi \qquad (4.15)$$

In this section we have focused on the subsonic strongly-interacting state. The formulation in similarity variables also gives us useful perspective for the supersonic cases, please see (Balsara [17], Vides, Nkonga & Audit [69], Balsara & Dumbser [18]). This completes our discussion of the multidimensional Riemann solver.

The fluxes in eqns. (4.14) and (4.15) can now be used to obtain properly upwinded electric fields at the edges of a Yee mesh. Notice from eqn. (2.1) that the negative of the seventh component of the x-flux as well as the sixth component of the y-flux will yield the z-component of the electric field. Now, both fluxes have the appropriate multidimensional contributions so that the best approach is to use $E_z = \left(-\left(\mathbf{F}_{numerical}\right)_7 + \left(\mathbf{G}_{numerical}\right)_6\right)/2$. (The subscripts "6" and "7" in the previous equation denote the sixth and seventh components of the y-flux and the x-flux respectively. The electric fields will be multidimensionally upwinded by the multidimensional Riemann solver described here and will avoid all the issues raised by Gardiner and Stone [43, 44].

The edge fluxes from the multidimensional Riemann solver can also be used to improve the facially-averaged fluxes for the conserved variables. Please see eqns. (24) and (25) from Balsara [12].

**V) Accuracy Analysis**

For classical hydrodynamics and MHD, there are several very nice, non-trivial test problems for demonstrating that a numerical method meets its design accuracy. For two dimensional hydrodynamical flows, a very interesting flow that propagates in a self-preserving fashion is provided by the isentropic vortex that is described in detail in Balsara & Shu [6]. For MHD, a corresponding magnetized vortex is presented in Balsara [9]. Three dimensional test problems that are suitable for hydrodynamical and MHD accuracy analysis are also presented in Balsara *et al*. [11]. For this work, we content ourselves with showing that very nice test problems can be designed for relativistic hydrodynamics and RMHD. The test problems presented here are the relativistic analogues of the classical hydrodynamical and MHD vortices.

The problem is set up on a periodic domain that spans $[-5,5] \times [-5,5]$. We first describe the velocity and magnetic field in the rest frame of the vortex. For nonrelativistic hydrodynamics or MHD, making the vortex move on the mesh is just a matter of adding a net velocity. For relativistic hydrodynamics and MHD, one has to include the additional complications of relativistic velocity addition and Lorentz transformation. These additional tasks are entirely non-trivial for relativistic flow. For that reason, we initially focus on the description of the vortex in its own rest frame. In a subsequent paragraph we will describe the velocity addition and Lorentz transformation. The velocity of the vortex (before it is made to move relative to the mesh) is given by

$$(v_x, v_y) = v_{max}^\phi e^{0.5(1-r^2)}(-y, x)$$

For both the hydrodynamical and RMHD test problems we have used $v_{max}^\phi = 0.7$. Notice that the velocity diminishes rapidly far away from the center of the vortex. This rapid drop in the velocity ensures that the boundaries of the domain have a negligible effect on the dynamics of the vortex. The magnetic field of the vortex (before it is made to move relative to the mesh) is given by

$$(B_x, B_y) = B_{max}^\phi e^{0.5(1-r^2)}(-y, x)$$

For the RMHD test problem we set $B_{max}^\phi = 0.7$. Notice that the magnetic field diminishes rapidly far away from the center of the vortex. This rapid drop in magnetic pressure and magnetic tension ensures that the boundaries of the domain have a negligible effect on the dynamics of the vortex. The corresponding magnetic vector potential, which is very useful for setting up a divergence-free vector field, is given by

$$A_z = B_{max}^\phi e^{0.5(1-r^2)}$$

The pseudo-entropy is defined by $S = P/\rho^\Gamma$ with polytropic index $\Gamma = 5/3$. The pressure and density of the vortex are also set to unity at the center of the vortex. The vortex is initialized to be isentropic so that $\delta S = 0$; i.e., the entropy is a constant throughout the vortex. Consistent with this velocity field and magnetic field, the steady state equation for the radial momentum of the vortex yields a pressure balance condition. This pressure balance condition for an RMHD vortex is given by

$$r\frac{d\mathrm{P}_{Tot}}{dr} = \left(\rho h + b^2\right)\gamma^2\left(\mathrm{v}^\phi\right)^2 - \left(b^\phi\right)^2$$

For the hydrodynamical case, the above equation simplifies to become.

$$r\frac{d\mathrm{P}_g}{dr} = \rho h \gamma^2 \left(\mathrm{v}^\phi\right)^2$$

Depending on the circumstance, one of the above two equations is numerically integrated radially outwards from the center of the vortex. Along with the isentropic condition, this equation fully specifies the run of the density and pressure in the vortex as a function of radius. Fig. 2a shows the run of thermal pressure as a function of radius for the vortices used here in the relativistic hydrodynamics and RMHD cases. Notice that a specification of the pressure at all radial points in the vortex also yields the density because of the isentropic condition. Observe that the thermal pressure profile for the magnetized vortex is less steep in Fig. 2a because the magnetic pressure supplements the gas pressure. This completes the description of the vortex in its own rest frame. Because the next steps associated with relativistic velocity addition and Lorentz transformation are non-trivial, we recommend that the run of density and pressure for the vortices should be tabulated on a very fine one-dimensional radial mesh. Typically, this radial mesh should have resolution that is much finer than the two-dimensional mesh on which the problem is computed.

We now describe the process of mapping the vortex to a computational mesh on which it moves with a speed $\beta_x \hat{x} + \beta_y \hat{y}$. We use $\beta_x = \beta_y = 0.5$ for our vortex; i.e. our vortex moves on the mesh with a speed that is $1/\sqrt{2}$ times the speed of light. Let us define $\gamma_\beta \equiv 1/\sqrt{1-\beta_x^2-\beta_y^2}$ to be the Lorentz factor associated with this velocity. In reality, this mapping of the vortex to a computational mesh is achieved by making the computational mesh move with a speed $-\beta_x \hat{x} - \beta_y \hat{y}$ relative to the rest frame of the vortex. Let the rest frame of the vortex be described by the unprimed spacetime coordinates $(t, x, y, z)^T$. The coordinates of the computational mesh, therefore, correspond to primed spacetime coordinates given by $(t', x', y', z')^T$. In practice, $t' = 0$ when initializing the computational mesh and realize too that the equations that describe the vortex in its own rest frame are also time-independent, i.e. they do not depend on "$t$". Thus for any chosen coordinate $(t' = 0, x', y', z' = 0)^T$ on the computational mesh we can find the corresponding unprimed coordinates via the following Lorentz transformation

$$\begin{pmatrix} t' \\ x' \\ y' \\ z' \end{pmatrix} = \begin{pmatrix} \gamma_\beta & \gamma_\beta \beta_x & \gamma_\beta \beta_y & 0 \\ \gamma_\beta \beta_x & 1+(\gamma_\beta-1)\beta_x^2/\beta^2 & (\gamma_\beta-1)\beta_x\beta_y/\beta^2 & 0 \\ \gamma_\beta \beta_y & (\gamma_\beta-1)\beta_x\beta_y/\beta^2 & 1+(\gamma_\beta-1)\beta_y^2/\beta^2 & 0 \\ 0 & 0 & 0 & 1 \end{pmatrix} \begin{pmatrix} t \\ x \\ y \\ z \end{pmatrix}$$

The unprimed coordinates refer to the rest frame of the vortex. In the unprimed frame, all the flow variables associated with the vortex have already been specified via the discussion in the previous paragraph. Scalar variables, like density and thermal pressure, are referred to the rest frame of the fluid parcel, i.e. they are proper variables that transform as scalars. Consequently, they transform unchanged as long as the Lorentz transform in the previous equation is properly applied. Three-velocities have to be suitably transformed using the relativistic addition of velocities. The appropriate formulae that give us the velocities in the primed frame from the original velocities in the unprimed frame are given below as:

$$v'_x = \frac{\gamma_\beta \beta_x + \left[1 + \frac{(\gamma_\beta-1)\beta_x^2}{\beta^2}\right] v_x + \frac{(\gamma_\beta-1)\beta_x\beta_y}{\beta^2} v_y}{\gamma_\beta (1 + \beta_x v_x + \beta_y v_x)}$$

and

$$v'_y = \frac{\gamma_\beta \beta_y + \frac{(\gamma_\beta-1)\beta_x\beta_y}{\beta^2} v_x + \left[1 + \frac{(\gamma_\beta-1)\beta_y^2}{\beta^2}\right] v_y}{\gamma_\beta (1 + \beta_x v_x + \beta_y v_x)}$$

With the relativistic velocity addition formulae described above, we can obtain the velocities at any point on our computational mesh. Since we use a magnetic vector potential to initialize our magnetic field, we point out that the electric field potential, $\Phi$, and the magnetic vector potential ($\vec{A}$) together form a four-vector $(\Phi, A^i)^T$. Being a four-vector, it transforms just like a four-coordinate. We can, therefore, obtain the magnetic vector potential in the primed frame. In the specific instance of the vortex that we describe here, the z-component of the magnetic vector potential is unchanged as we transform from the unprimed frame back to the primed frame. Even in the primed frame, the previously described Lorentz transformation is such that only the z-component of the magnetic vector potential will be non-zero. Likewise, the value of $\Phi$ is immaterial and set to zero. We see therefore that it is easy to initialize the divergence-free magnetic field for the vortex on the computational mesh. This completes our discussion of the set-up for relativistically boosted hydrodynamical and RMHD vortices on a computational mesh. Because these relativistic vortices are new in the literature, in Fig. 2b we show the density profile of the RMHD vortex on the computational mesh at the initial time. Notice that the boosted vortex shows substantial Lorentz contraction in its density variable. Figs. 2c and 2d show the x-velocity and y-velocity respectively. Notice that the velocity profiles are not symmetrical about the northeast-pointing diagonal of the mesh owing to the relativistic velocity addition formulae.

We have run the relativistic hydrodynamical vortex at several resolutions with our second order accurate ADER-WENO scheme. Table I shows the density and energy variable as a function of various resolutions. We see that our second order scheme meets its design accuracy. We have also run the RMHD vortex at several resolutions with our second order accurate ADER-WENO scheme. Table II shows the density and x-component of the magnetic field variable as a function of various resolutions. We see again that the code meets the design accuracy. We have also run the analogous simulations at third order (Table III and IV) and also at fourth order (Table V and VI). We see from those tables that the code agrees well with its design accuracy for all the ADER-WENO schemes that use the MuSIC Riemann solver.

**Table I) Accuracy analysis for second order relativistic hydrodynamical vortex simulation.**

| Zones | $L_1$ Error (density) | Order | $L_\infty$ Error (density) | Order | $L_1$ Error (energy) | Order | $L_\infty$ Error (energy) | Order |
|---|---|---|---|---|---|---|---|---|

| Zones | $L_1$ Error (density) | Order | $L_\infty$ Error (density) | Order | $L_1$ Error | Order | $L_\infty$ Error | Order |
|---|---|---|---|---|---|---|---|---|
| 64×64 | 7.41E-02 | | 2.90E+00 | | 1.39E+00 | | 4.65E+01 | |
| 128×128 | 1.59E-02 | 2.22 | 6.99E-01 | 2.05 | 3.22E-01 | 2.10 | 1.77E+01 | 1.39 |
| 256×256 | 3.37E-03 | 2.23 | 1.76E-01 | 1.99 | 7.08E-02 | 2.19 | 4.27E+00 | 2.05 |
| 512×512 | 7.47E-04 | 2.18 | 3.81E-02 | 2.21 | 1.64E-02 | 2.11 | 9.31E-01 | 2.20 |

**Table II) Accuracy analysis for second order RMHD vortex simulation.**

| Zones | $L_1$ Error (density) | Order | $L_\infty$ Error (density) | Order | $L_1$ Error ($B_x$) | Order | $L_\infty$ Error ($B_x$) | Order |
|---|---|---|---|---|---|---|---|---|
| 64×64 | 5.77E-02 | | 1.97E+00 | | 6.88E-03 | | 1.85E-01 | |
| 128×128 | 1.31E-02 | 2.14 | 5.64E-01 | 1.80 | 1.92E-03 | 1.84 | 6.54E-02 | 1.50 |
| 256×256 | 2.81E-03 | 2.22 | 1.40E-01 | 2.01 | 5.30E-04 | 1.85 | 1.92E-02 | 1.77 |
| 512×512 | 6.15E-04 | 2.19 | 3.03E-02 | 2.21 | 1.38E-04 | 1.95 | 4.71E-03 | 2.02 |

**Table III) Accuracy analysis for third order relativistic hydrodynamical vortex simulation.**

| Zones | $L_1$ Error (density) | Order | $L_\infty$ Error (density) | Order | $L_1$ Error (energy) | Order | $L_\infty$ Error (energy) | Order |
|---|---|---|---|---|---|---|---|---|
| 64×64 | 3.99E-02 | | 1.16E+00 | | 3.94E-03 | | 1.27E-01 | |
| 128×128 | 7.26E-03 | 2.46 | 2.68E-01 | 2.11 | 7.07E-04 | 2.48 | 2.91E-02 | 2.13 |
| 256×256 | 1.08E-03 | 2.76 | 5.09E-02 | 2.39 | 1.05E-04 | 2.74 | 5.71E-03 | 2.35 |
| 512×512 | 1.41E-04 | 2.93 | 7.31E-03 | 2.80 | 1.40E-05 | 2.91 | 8.14E-04 | 2.81 |

**Table IV) Accuracy analysis for third order RMHD vortex simulation.**

| Zones | $L_1$ Error (density) | Order | $L_\infty$ Error (density) | Order | $L_1$ Error ($B_x$) | Order | $L_\infty$ Error ($B_x$) | Order |
|---|---|---|---|---|---|---|---|---|
| 64×64 | 3.99E-02 | | 1.16E+00 | | 3.94E-03 | | 1.27E-01 | |
| 128×128 | 7.26E-03 | 2.46 | 2.68E-01 | 2.11 | 7.07E-04 | 2.48 | 2.91E-02 | 2.13 |
| 256×256 | 1.08E-03 | 2.76 | 5.09E-02 | 2.39 | 1.05E-04 | 2.74 | 5.71E-03 | 2.35 |

| 512×512 | 1.41E-04 | 2.93 | 7.31E-03 | 2.80 | 1.40E-05 | 2.91 | 8.14E-04 | 2.81 |

**Table V) Accuracy analysis for fourth order relativistic hydrodynamical vortex simulation**.

| Zones | $L_1$ Error (density) | Order | $L_\infty$ Error (density) | Order | $L_1$ Error (energy) | Order | $L_\infty$ Error (energy) | Order |
|---|---|---|---|---|---|---|---|---|
| 64×64 | 1.10E-02 |  | 4.02E-01 |  | 1.83E-01 |  | 7.88E+00 |  |
| 128×128 | 9.44E-04 | 3.54 | 5.42E-02 | 2.89 | 1.22E-02 | 3.91 | 1.08E+00 | 2.86 |
| 256×256 | 4.66E-05 | 4.34 | 3.39E-03 | 4.00 | 6.38E-04 | 4.26 | 7.90E-02 | 3.78 |
| 512×512 | 2.15E-06 | 4.44 | 1.58E-04 | 4.42 | 3.39E-05 | 4.23 | 4.12E-03 | 4.26 |

**Table VI) Accuracy analysis for fourth order RMHD vortex simulation**.

| Zones | $L_1$ Error (density) | Order | $L_\infty$ Error (density) | Order | $L_1$ Error ($B_x$) | Order | $L_\infty$ Error ($B_x$) | Order |
|---|---|---|---|---|---|---|---|---|
| 64×64 | 8.69E-03 |  | 3.21E-01 |  | 8.72E-04 |  | 4.02E-02 |  |
| 128×128 | 7.59E-04 | 3.52 | 4.01E-02 | 3.00 | 8.98E-05 | 3.28 | 6.56E-03 | 2.62 |
| 256×256 | 3.70E-05 | 4.36 | 2.55E-03 | 3.97 | 6.66E-06 | 3.75 | 5.44E-04 | 3.59 |
| 512×512 | 1.63E-06 | 4.51 | 1.13E-04 | 4.49 | 4.46E-07 | 3.90 | 3.47E-05 | 3.97 |

## VI) Test Problems

We show that our method performs well for a range of stringent test problems that are presented in this Section. Some of the tests are well-known; whereas the tests in Sub-sections VI.c and VI.e are genuinely novel to RMHD simulations. While the tests in this Section are shown at specific orders of accuracy, we have also validated most of them at all the orders of accuracy that have been presented in this paper.

## VI.a) RMHD Rotor Problem

The rotor test problem was initially presented for classical MHD by Balsara & Spicer [5] and it has been adapted to RMHD by Del Zanna *et al*. [31] in two-dimensions and Mignone *et.*

*al*. [57] in three-dimensions. We find that there are nuances to this problem when it is extended to RMHD and we point out that those nuances have gone unappreciated in the RMHD literature. We focus on the two-dimensional version of this problem. The problem is set up on a unit domain in two dimensions which spans $[-0.5, 0.5] \times [-0.5, 0.5]$. A unit x-magnetic field is set up all over the domain with a unit thermal pressure. There is a unit density in the problem everywhere except within a radius of 0.1, where the density becomes ten times larger. The high density region is set into rapid rotation with a velocity given by $\vec{v}(x,y) = -w\,y\,\hat{x} + w\,x\,\hat{y}$, thus forming a rotor. The parameter "$w$" controls the rotation speed. For example, $w = 9.95$ would give us a Lorentz factor of 10 at a radius of 0.1; though higher values of "$w$" can certainly be chosen for setting up more stringent versions of this problem. A taper is usually applied to the rotor's rotational velocity so that it smoothly joins the non-rotating ambient medium within a radial distance of six mesh zones.

Because very small changes in "$w$" can result in very large changes in the Lorentz factor, the problems arise when one tries to set up this problem on a computational mesh. The high Lorentz factor flows are confined to a very thin ring at the outer boundary of the rotor. Traditionally, one initializes the variables at the zone-center of each zone and we have followed that trend because it is certainly valid for a second order scheme. We set up the problem on uniform meshes of various sizes ranging from $200 \times 200$ zones all the way up to $6000 \times 6000$ zones. (Traditionally, this problem has been reported on a $400 \times 400$ zone mesh.) We then evaluated the fraction of zones that are within 10% and 20% of the targeted maximal Lorentz factor in the problem for that particular size of mesh. The traditional value of $w = 9.95$, which yields a maximal Lorentz factor of 10 is shown in Fig. 3a. We see from Fig. 3a that none of the zones on a $400 \times 400$ zone mesh will have a Lorentz factor that is within 10% of the maximal value. We also see from Fig. 3a that as the number of zones is increased, the fraction of zones that have Lorentz factor that is within 10% of the maximal value increases. At some resolution this fraction becomes asymptotically flat in Fig. 3a. However, please realize that this does not happen till we reach meshes with at least $1200 \times 1200$ zones. Fig. 3b provides information that is analogous to Fig. 3a, with the exception that we used $w = 9.9944$ which corresponds to a maximal Lorentz factor of 30. Please notice from Fig. 3b that the fraction of zones that lie within 10% of a maximal Lorentz factor of 30 does not become asymptotically flat till the mesh has at least $3500 \times 3500$ zones. In fact, we see that a $400 \times 400$ zone mesh would not have any zones that reach the desired Lorentz factor! This proves that in order to do this test problem with fidelity, we should go to very high resolution meshes.

We first used $w = 9.95$, which corresponds to a maximal Lorentz factor of 10, on a $2500 \times 2500$ zone mesh. (Somewhat lower resolutions can, of course, be used for this Lorentz factor of 10 simulation based on guidance obtained from Fig. 3a.) Figs. 4a through 4d show the density, gas pressure, Lorentz factor and magnetic field magnitude at a final time of 0.4. We then used $w = 9.9944$, which corresponds to a maximal Lorentz factor of 30, on a $4700 \times 4700$ zone mesh. Figs. 5a through 5d show the density, gas pressure, Lorentz factor and magnetic field magnitude at a final time of 0.4. Despite the very large initial Lorentz factor, we see that all the flow variables are well-represented. A cursory viewing might suggest that Fig. 4 is not very different from Fig. 5, but closer inspection shows important differences. Realize that the density and gas pressure are significantly higher in Figs. 5a and 5b than the analogous variables in Fig. 4a and 4b. Furthermore, the flow with the larger Lorentz factor produces a greater outward expansion in the density owing to its larger centrifugal effect. The final Lorentz factor in Fig. 5c is also significantly higher than the final Lorentz factor in Fig. 4c. The larger simulation also shows a hint of a Kelvin-Helmholtz instability, though we did not pursue that further in this paper. Similarly, the magnetic field in Fig. 5d is much more compressed compared to Fig. 4d. The simulations in Figs. 4 and 5 were run with a CFL of 0.4 using a third order accurate ADER-WENO scheme along with the MuSIC Riemann solver.

**VI.b) RMHD Blast Problem**

The non-relativistic version of this test problem was first presented in Balsara & Spicer [5] and was extended to RMHD by Komissarov [50]. Several variants of this test problem have been presented by Mignone & Bodo [56] and Del Zanna et al. [32] by using a slightly modified version of the moderately magnetized case in Komissarov [50]. In this subsection, we use the problem setup by Mignone & Bodo [56] and Del Zanna et al. [32].

The problem is set up on a unit domain in two dimensions which spans $[-6, 6] \times [-6, 6]$ in the xy-plane. We used a mesh with $400 \times 400$ zones. Within a radius of 0.8, the explosion zone has a density of $10^{-2}$ and a pressure of 1. Outside a radius of 1, the ambient medium has a density of $10^{-4}$ and a pressure of $5\times 10^{-4}$. A linear taper is applied to the density and pressure from a radius of 0.8 to 1. Accordingly, both the density and pressure linearly decrease with increasing radius in that range of radii. The magnetic field is initialized in the x-direction and has a magnitude of 0.1. The polytropic index of $\Gamma = 4/3$ is used in this problem. The simulation is run to a final time of 4 with a CFL of 0.5 using a fourth order accurate scheme. Please notice that this is a problem with near-infinite shocks, even so, it runs stably with a fourth order accurate ADER-WENO scheme with the MuSIC Riemann solver that is based on the methods developed here.

Figs. 6a and 6b show the $\log_{10}$ of the density and gas pressure. Figs. 6c and 6d show the Lorentz factor and the magnitude of the magnetic field. The third order accurate scheme was used in this simulation. We see that the RMHD result is properly reproduced.

**VI.c) RMHD Vortex Interacting with a Shock**

Vorticity is known to be a very important dynamical variable in the evolution of hydrodynamical turbulence. Vortices are also known to be amazingly stable fluid dynamical structures. When hydrodynamical vortices interact with shocks of mild to modest strength, they retain their structure. This was first shown computationally by Pao and Salas [60]. These results were later amplified by Meadows, Kumar and Hussaini [53], Jiang and Shu [48] and Balsara and Shu [6]. There is a lot of recent interest in RMHD turbulence. It is, therefore, interesting to demonstrate the stability of RMHD vortices as they interact with relativistic shocks. Apart from being an interesting demonstration that the insights from non-relativistic hydrodynamics extend also to RMHD, the problem described here also makes for an interesting test problem.

This RMHD shock-vortex interaction problem uses the same vortex parameters in the rest frame of the vortex as were used in Section V. However, the boosted velocity of the RMHD vortex is now set to $\beta_x \hat{x} + \beta_y \hat{y} = 0.6\hat{x} + 0.6\hat{y}$, i.e. the vortex moves slightly faster. The computational domain spans $[-6,6]\times[-6,6]$ and the vortex is initially centered at $(-3,-3)$. The ambient fluid around the vortex moves with the boosted velocity of the vortex. The RMHD vortex and its ambient fluid are initially unshocked. Sufficiently far away from the center of the vortex, the magnetic field is almost zero. Part of the computational domain with $y \geq -x-2$ has higher density and pressure with a lower velocity. I.e. the line corresponding to $y = -x - 2$ has a standing shock. The density, pressure and velocity in the post-shock region are given by

$$\rho = 10.47090373 \quad ; \quad P_g = 50.44557978 \quad ; \quad v_x = v_y = 0.507707117$$

These post-shock parameters are so designed that the shock is a standing shock on the computational mesh. The boundary conditions retain the shock profile so that there is practically no back-reaction from the boundaries into the computational domain. The density and the pressure in the unshocked ambient medium around the vortex should be computed by solving for the pressure equation, as described in Section V. Even so, it may be useful to give the values in the unshocked gas far away from the center of the vortex. They are given by

$$\rho = 6.73586072 \quad ; \quad P_g = 24.02454458 \quad ; \quad v_x = v_y = 0.6$$

The entire domain has an ideal gas with a polytropic index of $\Gamma = 5/3$, identical to Section V. The entire simulation is run to a time of $t = 10$ by which point the vortex has crossed the shock and now resides in the post-shock region. This completes our description of the problem set-up.

The results of the RMHD shock-vortex interaction problem are shown in Fig. 7. A third order scheme with a CFL of 0.45 was used on a mesh with $500 \times 500$ zones. Our scheme used the ADER-WENO predictor along with the MuSIC Riemann solver-based corrector. Figs. 7a, 7b and 7c show the density, Lorentz factor and magnetic field magnitude in the simulation at time $t = 0$, i.e. the initial set-up. Figs. 7d, 7e and 7f show the density, Lorentz factor and magnetic field magnitude in the simulation at time $t = 3.4$, i.e. when the vortex is half-way through the shock. Figs. 7g, 7h and 7i show the density, Lorentz factor and magnetic field magnitude in the simulation at time $t = 10$, i.e. at the final time of the simulation. We see that when the vortex is crossing the shock, it emits several sound waves. We also see that after the vortex has managed to pass through the shock, its form is preserved and it is convected with the post-shock velocity. I.e. the RMHD vortex is stable after interaction with the shock. Since the final post-shock velocity is lower than the velocity of the unshocked gas, the vortex shows a rounder profile in the post-shock region. I.e. the effects of Lorentz contraction are reduced when the vortex reaches the post-shock region. Our simulation confirms our expectation that an RMHD vortex can stably interact with shocks of modest strength.

## VI.d) RMHD Jet Problem

The propagation of a relativistic axisymmetric jet in cylindrical coordinates (r, z) is presented in this sub-section. We follow the details of the setup by Mignone & Bodo [56]. The problem is initialized on a unit domain in two dimensions which spans $[0,12] \times [0,50]$ in $rz$-plane. We used a mesh with $120 \times 500$ zones. Variables with a subscript of "*j*" correspond to jet values; variables with a subscript of "*a*" correspond to ambient values. An ambient medium that has uniform density, pressure and magnetic field is initialized with following values:

$$\rho_a = 1, \quad P_a = \frac{\eta v^2}{\Gamma(\Gamma-1)M^2 - \Gamma v^2}, \quad \mathbf{B}_a = \sqrt{2P_a}\hat{z}.$$

Here, Mach number of jet, $M = v_j/c_j = 6$, density ratio between jet and ambient medium, $\eta = \rho_j/\rho_a = 0.01$, speed of jet, $v_j = 0.99$ and polytropic index, $\Gamma = 5/3$ are used in this simulation. We use a reflective boundary condition at the lower *z* boundary except for the jet nozzle ($r < 1$), which is treated with an inflow boundary condition. The jet nozzle has a constant inflow with time where the inflow parameters are $\rho_j = 0.01$, $P_j = P_a$, $\mathbf{v} = v_j\hat{z}$ and $\mathbf{B}_j = \sqrt{2P_a}\hat{z}$. The upper *r*- and *z*-boundaries were continuative outflow. The simulation is run to a time of $t = 126$ with CFL of 0.3. A second order of accurate scheme based on methods developed here was used in this simulation.

The results of the propagation of axisymmetric relativistic jet problem are shown in Fig. 8. Figs. 8a, 8b and 8c show the density, magnetic pressure in log scale and Lorentz factor in the simulation at final time. Notice that the Kelvin-Helmholtz instability is observed due to the velocity shear between jet and ambient medium.

## VI.e) Long-Term Decay of Alfven Waves

Turbulence in non-relativistic and relativistic plasmas is currently one of the hot topics in astrophysics. We know that the turbulence in magnetized plasmas is Alfvenic; i.e., the propagation and interaction of Alfven waves gives rise to turbulence. In order for RMHD turbulence to be correctly represented, we need to ensure that isolated, torsional Alfven waves can propagate with minimal numerical dissipation on a computational mesh. The RMHD wave families can propagate at 45° to the mesh lines with minimum dissipation. It is much more difficult to achieve good propagation of waves that are required to propagate at a small angle to one of the mesh lines (Balsara [9]).

We construct an RMHD version of a test problem that examines the dissipation of torsional Alfven waves when they propagate at a small angle to the mesh. See Balsara [9] for non-relativistic test. We use a uniform $120 \times 120$ zone mesh that spans $[-3,3] \times [-3,3]$ in the xy-plane. An uniform density, $\rho_0 = 1$, and pressure, $P_0 = 1$, are initialized on the mesh. The unperturbed velocity is $v_0 = 0$, and the unperturbed magnetic field is $\mathbf{B}_0 = 0.5$. A polytropic index of $\Gamma = 4/3$ is used in this simulation. The amplitude of the Alfven wave fluctuation ($B_1$) can be parameterized in terms of the velocity fluctuation, which has a value of 0.1 in this problem. The Alfven wave is designed to propagate along the wave vector, $\mathbf{k} = k_x \hat{x} + k_y \hat{y}$, where $k_x = 1/6$, $k_y = 1$. The velocity and magnetic field are given as follows:

$$\mathbf{v} = v_1 n_y \cos\phi \hat{x} - v_1 n_x \cos\phi \hat{y} + v_1 \sin\phi \hat{z} ,$$

$$\mathbf{B} = [B_0 n_x + B_1 n_y \cos\phi]\hat{x} + [B_0 n_y - B_1 n_x \cos\phi]\hat{y} + B_1 \sin\phi \hat{z} .$$

Here, the unit vector, $\mathbf{n} = n_x \hat{x} + n_y \hat{y} = (k_x \hat{x} + k_y \hat{y})/\sqrt{k_x^2 + k_y^2}$, the phase of the wave at initial time, $\phi = 2\pi(k_x x + k_y y)$, and the perturbation amplitude of the magnetic field is given by $B_1 = v_1 \sqrt{\rho_0 + \frac{\Gamma}{\Gamma - 1} P_0 + B_0^2}$. The corresponding vector potential for the magnetic field is given by

$$\mathbf{A} = \frac{B_1}{2\pi k_x} \cos\phi \hat{y} + \left[ B_0 (n_x y - n_y x) - \frac{B_1}{2\pi \sqrt{k_x^2 + k_y^2}} \sin\phi \right] \hat{z} .$$

The entire simulation is run to a time of $t = 130$ by which time the Alfven waves have crossed the computational domain five times. A CFL of 0.4 is used.

The maximum z-component of velocity and magnetic field in log scale as a function of time with various order of accuracy are shown in Fig 9a and Fig9b, respectively. The second, third and fourth order schemes from this paper are applied to the simulation. The amount of dissipation decreases as the order of accuracy increases and almost no dissipation is observed in the fourth order simulation.

## VII) Conclusions

In this paper we report on three advances in numerical RMHD. The first advance, which admittedly was known before, is that there is a uniquely good reconstruction strategy for RMHD. The reconstruction strategy consists of reconstructing neither on the conserved variables nor on the primitive variables which include the velocity. Instead, one should reconstruct on the primitive variables formed by the product of the Lorentz factor and the velocity. This always results in a velocity that is subluminal, especially at large Lorentz factors. The important new innovation in this paper consists of showing that the reconstruction in primitive variables can even be carried out with higher order accuracy. The second advance, which is novel, consists of a predictor step based on the primitive variables consisting of the product of the Lorentz factor and the velocity. This combination ensures that the predictor step will keep the four velocity subluminal. The third advance, which is also genuinely novel, consists of designing a multidimensional Riemann solver for RMHD. This MuSIC Riemann solver enables us to get highly accurate, multidimensionally upwinded electric fields at the edge-centers of the mesh. This, in turn, improves the propagation of the face-centered magnetic field.

The methods presented in this paper are very general and can also be applied to other PDE systems where physical realizability is most easily asserted in the primitive variables. We show that primitive variables can also be used for higher order schemes, including schemes that go well beyond second order. This is a very novel demonstration because prior experience had suggested that in order to go beyond second order accuracy one absolutely needed to reconstruct the conserved variables. Very interestingly, the standard WENO reconstruction methods and ADER time-update methods can be adapted with minimal changes even for very high order schemes.

We also provide a very compact description of the multidimensional Riemann solver. Our compact description of the multidimensional Riemann solver is intended to facilitate implementation, although we do urge the reader to read Balsara *et al*. [23] for a thorough description of the underlying theory. Such MuSIC Riemann solvers have been shown to provide greater physical fidelity to all manner of CFD simulations (Balsara [14], [17]). They have great utility for a variety of applications like ALE simulations (Boscheri, Balsara and Dumbser [26], Boscheri, Dumbser and Balsara [27]) where physics-based node motion is especially difficult to achieve. Their great utility for involution-constrained systems has also been recently brought to the forefront for the MHD system (Balsara [14], [17], Balsara and Dumbser [19]) and for Maxwell's equations (Balsara *et al*. [22]). RMHD also constitutes an involution-constrained system and the present paper shows the value of multidimensional Riemann solvers for this important hyperbolic system.

We also design stable vortices for relativistic hydrodynamics and RMHD. These serve as useful test problems for proving that RMHD codes indeed meet their design accuracy. Several other stringent test problems are also presented. The importance of resolution for genuinely treating the magnetized RMHD rotor problem is also shown. Our test problems also include a novel demonstration that RMHD vortices are stable when they interact with shocks.

**Acknowledgements**

DSB acknowledges support via NSF grants NSF-ACI-1307369, NSF-DMS-1361197 and NSF-ACI-1533850. DSB also acknowledges support via NASA grants from the Fermi program as well as NASA-NNX 12A088G. Several simulations were performed on a cluster at UND that is run by the Center for Research Computing.  Computer support on NSF's XSEDE computing resources is also acknowledged.

**Appendix A**

In this appendix, we present components of characteristic matrices shown in Eq. (3.1). Only characteristic matrix **A** that corresponds to x-flux is shown here. Others such as **B** and **C** matrices can be easily obtained by considering symmetries of the equations. The Jacobian matrix $\partial \mathbf{U}/\partial \mathbf{V}$ is

$$\left(\frac{\partial \mathbf{U}}{\partial \mathbf{V}}\right) = \begin{bmatrix} \gamma & \rho v_x & \rho v_y & \rho v_z & 0 & 0 & 0 & 0 \\ \gamma^2 v_x & & & & Cv_x & & & \\ \gamma^2 v_y & & \Pi_{ij} & & Cv_y & & \Theta_{ij} & \\ \gamma^2 v_z & & & & Cv_z & & & \\ \gamma^2 & & \Sigma_i & & C-1 & & Z_i & \\ 0 & 0 & 0 & 0 & 0 & 1 & 0 & 0 \\ 0 & 0 & 0 & 0 & 0 & 0 & 1 & 0 \\ 0 & 0 & 0 & 0 & 0 & 0 & 0 & 1 \end{bmatrix},$$

where $C = \Gamma \gamma^2 / (\Gamma - 1)$.

$\Pi_{ij}$, $\Theta_{ij}$, $\Sigma_i$ and $Z_i$ are as follows:

$$\Pi_{ij} = \frac{1}{\gamma} \left[ \rho h \gamma^2 (v_i v_j + \delta_{ij}) - B^2 (v_i v_j - \delta_{ij}) - B_i B_j + (\mathbf{v} \cdot \mathbf{B}) B_i v_j \right],$$

$$\Theta_{ij} = 2 v_i B_j - B_i v_j - (\mathbf{v} \cdot \mathbf{B}) \delta_{ij},$$

$$\Sigma_i = \frac{1}{\gamma} \left[ \left(2 \rho h \gamma^2 + b^2\right) v_i - (\mathbf{v} \cdot \mathbf{B}) B_i \right],$$

$$Z_i = \left(1 + v^2\right) B_i - (\mathbf{v} \cdot \mathbf{B}) v_i,$$

where $\delta_{ij}$ is Kronecker delta.

The matrix $\partial \mathbf{F}/\partial \mathbf{V}$ is

$$\left(\frac{\partial \mathbf{F}}{\partial \mathbf{V}}\right) = \begin{bmatrix} u_x & \rho & 0 & 0 & 0 & 0 & 0 & 0 \\ u_x^2 & A_{22} & A_{23} & A_{24} & Cv_x^2+1 & A_{26} & A_{27} & A_{28} \\ u_x u_y & A_{32} & A_{33} & A_{34} & Cv_x v_y & A_{36} & A_{37} & A_{38} \\ u_x u_z & A_{42} & A_{43} & A_{44} & Cv_x v_z & A_{46} & A_{47} & A_{48} \\ \gamma^2 v_x & & \Pi_{1j} & & Cv_x & & \Theta_{1j} & \\ 0 & 0 & 0 & 0 & 0 & 0 & 0 & 0 \\ 0 & A_{72} & A_{73} & A_{74} & 0 & -v_y & v_x & 0 \\ 0 & A_{82} & A_{83} & A_{84} & 0 & -v_z & 0 & v_x \end{bmatrix}.$$

Notice that the fifth low of $(\partial \mathbf{F}/\partial \mathbf{V})$ is equivalent to the second low of $(\partial \mathbf{U}/\partial \mathbf{V})$. Matrix components denoted by $A_{ij}$ are as follows.

$$A_{22} = \left[\left[2\left(\rho h\gamma^2 + B^2\left(1-v_x^2\right)+(\mathbf{v}\cdot\mathbf{B})B_x v_x\right)-b^2\right]u_x + \left[2(v_x b_x - u_x B_x)-b_0\right]B_x\right]/\gamma^2$$

$$A_{23} = \left[2\left((\mathbf{v}\cdot\mathbf{B})B_x v_y - B^2 v_x v_y - B_x B_y\right)u_x + 2b_x B_x v_y + b_0 B_y - b^2 u_y\right]/\gamma^2$$

$$A_{24} = \left[2\left((\mathbf{v}\cdot\mathbf{B})B_x v_z - B^2 v_x v_z - B_x B_z\right)u_x + 2b_x B_x v_z + b_0 B_z - b^2 u_z\right]/\gamma^2$$

$$A_{32} = \left[2v_x^2\left[(\mathbf{v}\cdot\mathbf{B})B_y - B^2 v_y\right] + m_y + 2b_y B_x v_x/\gamma - B_x\left(B_x v_y + B_y v_x\right)\right]/\gamma$$

$$A_{33} = \left[v_x\left(\rho h\gamma^2 + B^2 - B_y^2\right) + 2v_x v_y\left((\mathbf{v}\cdot\mathbf{B})B_y - B^2 v_y\right) + 2b_y B_x v_y/\gamma - B_x\left((\mathbf{v}\cdot\mathbf{B})+B_y v_y\right)\right]/\gamma$$

$$A_{34} = \left[2v_x v_z\left[(\mathbf{v}\cdot\mathbf{B})B_y - B^2 v_y\right] + 2b_y B_x v_z/\gamma - B_z\left(B_x v_y + B_y v_x\right)\right]/\gamma$$

$$A_{42} = \left[2v_x^2\left[(\mathbf{v}\cdot\mathbf{B})B_z - B^2 v_z\right] + m_z + 2b_z B_x v_x/\gamma - B_x\left(B_x v_z + B_z v_x\right)\right]/\gamma$$

$$A_{43} = \left[2v_x v_y\left[(\mathbf{v}\cdot\mathbf{B})B_z - B^2 v_z\right] + 2b_z B_x v_y/\gamma - B_y\left(B_x v_z + B_z v_x\right)\right]/\gamma$$

$$A_{44} = \left[v_x\left(\rho h\gamma^2 + B^2 - B_z^2\right) + 2v_x v_z\left((\mathbf{v}\cdot\mathbf{B})B_z - B^2 v_z\right) + 2b_z B_x v_z/\gamma - B_x\left((\mathbf{v}\cdot\mathbf{B})+B_z v_z\right)\right]/\gamma$$

$$A_{26} = -\left(B_y v_y + B_z v_z\right)v_x - \left(1/\gamma^2 + v_x^2\right)B_x,$$

$$A_{27} = 2(B_y v_x - B_x v_y) v_x - b_x/\gamma,$$

$$A_{28} = 2(B_z v_x - B_x v_z) v_x - b_z/\gamma,$$

$$A_{36} = 2(B_x v_y - B_y v_x) v_x - b_y/\gamma,$$

$$A_{37} = 2(B_y v_x - B_x v_y) v_y - b_x/\gamma,$$

$$A_{38} = (B_z v_y - B_y v_z) v_x + (B_z v_x - B_x v_z) v_y,$$

$$A_{46} = 2(B_x v_z - B_z v_x) v_x - b_z/\gamma,$$

$$A_{47} = (B_y v_z - B_z v_y) v_x + (B_y v_x - B_x v_y) v_z,$$

$$A_{48} = 2(B_z v_x - B_x v_z) v_z - b_x/\gamma,$$

$$A_{72} = \frac{1}{\gamma}\left[(B_x v_y - B_y v_x) v_x + B_y\right],$$

$$A_{73} = \frac{1}{\gamma}\left[(B_x v_y - B_y v_x) v_y + B_x\right],$$

$$A_{74} = \frac{v_z}{\gamma}(B_x v_y - B_y v_x),$$

$$A_{82} = \frac{1}{\gamma}\left[(B_x v_z - B_z v_x) v_x + B_z\right],$$

$$A_{83} = \frac{v_y}{\gamma}(B_x v_z - B_z v_x),$$

$$A_{84} = \frac{1}{\gamma}\left[(B_x v_z - B_z v_x) v_z + B_x\right].$$

$u_x$, $u_y$ and $u_z$ in the above matrix represent special components of 4-velocities i.e. $u_i = \gamma v_i$.

**Appendix B**

Kim & Balsara [49] presented the HLLC Riemann solver by expanding the Newtonian work of Honkkila & Janhunen [47] to RMHD. In our previous work, we provided jumps in the transverse velocity in addition to density jumps. However, the longitudinal velocity in the two resolved states as well as the pressure in those states did not fully conform to the suggestion of Toro, Spruce and Speares [68]. In this Appendix we rectify that deficiency.

Our notation in this appendix follows that in Kim & Balsara [49]. We denote the right and left states input with a subscript "R" and "L". The input states have no superscript. The right-going and left-going extremal waves in the HLLC Riemann solver have speeds $\lambda_R$ and $\lambda_L$ respectively. The HLLC Riemann solver puts two resolved states between the input left and right states. The two resolved states are separated by a contact distontinuity with propagates with the longitudinal velocity ($v_x^*$) that is shared by the two resolved states. Likewise, the total pressure in the two resolved states is denoted by ($P_{Tot}^*$) and remains constant across the contact discontinuity. The remaining variables in the resolved states are denotes with a subscript of "R" or "L" depending on right or left resolved state; but they also have a superscript star to denote that they are resolved states. Following Toro, Spruce and Speares [68] we wish to keep the longitudinal velocity and total pressure constant across both the resolved states that arise in the HLLC Riemann solver. It makes physical sense since the two quantities are continuous across the entropy wave.

To obtain the fully consistent values in the intermediate state, we again start with the Rankine-Hugoniot jump condition across the right-going wave:

$$\rho_R^* \gamma_R^* \left( v_x^* - \lambda_R \right) = \rho_R \gamma_R \left( v_{xR} - \lambda_R \right) \tag{A1}$$

$$m_{xR}^* v_x^* - B_x \frac{b_{xR}^*}{\gamma_R^*} + P_{Tot}^* - \lambda_R m_{xR}^* = m_{xR} v_{xR} - B_x \frac{b_{xR}}{\gamma_R} + P_{Tot,R} - \lambda_R m_{xR} \tag{A2}$$

$$m_{yR}^* v_x^* - B_x \frac{b_{yR}^*}{\gamma_R^*} - \lambda_R m_{yR}^* = m_{yR} v_{xR} - B_x \frac{b_{yR}}{\gamma_R} - \lambda_R m_{yR} \tag{A3}$$

$$m_{zR}^* v_x^* - B_x \frac{b_{zR}^*}{\gamma_R^*} - \lambda_R m_{zR}^* = m_{zR} v_{xR} - B_x \frac{b_{zR}}{\gamma_R} - \lambda_R m_{zR}, \tag{A4}$$

$$m_{xR}^* - \lambda_R \mathcal{E}_R^* = m_{xR} - \lambda_R \mathcal{E}_R, \tag{A5}$$

In eqns. (A2) to (A5), we first need to replace $m_{xR}^*$, $m_{yR}^*$ and $m_{zR}^*$ by the relation between momentum density and velocity. We then eliminate $\mathcal{E}_R^*$ from eqn. (A2) by multiplying eqn. (A5) with $v_{xR}^*$ and subtracting the result from eqn. (A2). A similar operation can be done for eqns. (A3) and (A4). Consequently, we get our final form of the equations for longitudinal velocity and total pressure:

$$\left(1 - \lambda_R v_x^*\right) P_{Tot}^* - B_x^2 \left[1 - \left(v_R^*\right)^2\right] + \left(v_R^* \cdot B_R^*\right) B_x \left(\lambda_R - v_x^*\right) + \left(m_{xR} - \lambda_R \mathcal{E}_R\right) v_x^* - m_{xR} v_{xR} + B_x \frac{b_{xR}}{\gamma_R} - P_{Tot,R} + \lambda_R m_{xR} = 0$$
$$\tag{A6}$$

as well as transverse velocities:

$$\left(\lambda_R \mathcal{E}_R - m_{xR}\right) v_{yR}^* + P_{Tot}^* v_{yR}^* \lambda_R - B_{yR}^* \left(v_R^* \cdot B_R^*\right) \left(\lambda_R - v_x^*\right) + B_x B_{yR}^* \left[1 - \left(v_R^*\right)^2\right] - B_x \frac{b_{yR}}{\gamma_R} - m_{yR} \left(\lambda_R - v_{xR}\right) = 0$$
$$\tag{A7}$$

$$\left(\lambda_R \mathcal{E}_R - m_{xR}\right) v_{zR}^* + P_{Tot}^* v_{zR}^* \lambda_R - B_{zR}^* \left(v_R^* \cdot B_R^*\right) \left(\lambda_R - v_x^*\right) + B_x B_{zR}^* \left[1 - \left(v_R^*\right)^2\right] - B_x \frac{b_{zR}}{\gamma_R} - m_{zR} \left(\lambda_R - v_{xR}\right) = 0$$
$$\tag{A8}$$

We focus only on the right state, shown with a subscript "$R$". We note that the subscript "$R$" can be replaced with "$L$" in order to get the left state in eqns. (A1) – (A8).

We solve eqns. (A6), (A7) and (A8) using a Newton-Raphson method. Notice that the Lorentz factor in the starred state ($1/\sqrt{1 - \left(v_R^*\right)^2}$) does not appear in eqns. (A6) - (A8) with the result that the iteration process is stable. If we refer to the LHS of eqns. (A7) and (A8) as $F_1$ and $F_2$, respectively, the corrections of transverse velocities ($\delta v_{yR}^*$ and $\delta v_{zR}^*$) are obtained by the inversion of the following 2×2 matrix relation:

$$\begin{pmatrix} a_{11} & a_{12} \\ a_{21} & a_{22} \end{pmatrix} \begin{pmatrix} \delta v^*_{yR} \\ \delta v^*_{zR} \end{pmatrix} = \begin{pmatrix} F_1 \\ F_2 \end{pmatrix}, \tag{A9}$$

where the matrix coefficients in eqn. (12) are

$$a_{11} = m_{xR} - \lambda_R \mathcal{E}_R - P^*_{Tot}\lambda_R + \left(B^*_{yR}\right)^2 \left(\lambda_R - v^*_x\right) + 2B_x B^*_{yR} v^*_{yR} \ ; \ a_{12} = B^*_{yR} B^*_{zR}\left(\lambda_R - v^*_x\right) + 2B_x B^*_{yR} v^*_{zR} \ ;$$

$$a_{21} = B^*_{yR} B^*_{zR}\left(\lambda_R - v^*_x\right) + 2B_x B^*_{zR} v^*_{yR} \ ; \ a_{22} = m_{xR} - \lambda_R \mathcal{E}_R - P^*_{Tot}\lambda_R + \left(B^*_{zR}\right)^2 \left(\lambda_R - v^*_x\right) + 2B_x B^*_{zR} v^*_{zR}.$$

Likewise, $G_L$ and $G_R$ are denoted by the LHS of eqn. (A8) for the left and right states, respectively. Then the corrections of longitudinal velocity and total pressure ($\delta v^*_x$ and $\delta P^*_{Tot}$) are obtained by the inversion of the following 2×2 matrix relation:

$$\begin{pmatrix} b_{11} & b_{12} \\ b_{21} & b_{22} \end{pmatrix} \begin{pmatrix} \delta v^*_x \\ \delta P^*_{Tot} \end{pmatrix} = \begin{pmatrix} G_R \\ G_L \end{pmatrix}, \tag{A10}$$

where the coefficients are

$$b_{11} = -\lambda_R P^*_{Tot} + B^2_x\left(\lambda_R + v^*_x\right) - \left(v^*_R \cdot B^*_R\right)B_x + m_{xR} - \lambda_R \mathcal{E}_R \ ; \ b_{12} = 1 - \lambda_R v^*_x;$$

$$b_{21} = -\lambda_L P^*_{Tot} + B^2_x\left(\lambda_L + v^*_x\right) - \left(v^*_L \cdot B^*_L\right)B_x + m_{xL} - \lambda_L \mathcal{E}_L \ ; \ b_{22} = 1 - \lambda_L v^*_x.$$

Iterations for the longitudinal velocity ($v^*_x$) and total pressure ($P^*_{Tot}$) in addition to the transverse velocities ($v^*_{yR,L}$ and $v^*_{zR,L}$) occur until solutions converge. After obtaining the velocities and total pressure in the intermediate states, we can obtain density and energy jumps from eqns. (A1) and (A5). As a result, we can evaluate all the values fully consistent with all jump conditions.

**Figure Captions**

*Fig 1a shows the initial situation where the four states come together at a vertex (shown by the gray dot). In time, the states evolve and we schematically show the wave model of an HLLC Riemann solver with two extremal waves and one central wave. In Fig. 1b we show four one-dimensional Riemann problems which interact with one-another to produce a strongly interacting state (S.I. State); see the thick dashed lines. The extremal speeds associated with the*

*S.I. State are also shown. In Fig. 1b, the strongly interacting state itself provides the resolved state at the vertex.*

*Fig. 2a shows the run of thermal pressure as a function of radius for the vortices used here in the relativistic hydrodynamics and RMHD cases. Fig. 2b we show the density profile of the RMHD vortex on the computational mesh at the initial time. Figs. 2c and 2d show the x-velocity and y-velocity respectively.*

*Fig. 3 plots the fraction of zones on the mesh for the rotor problem that are within 10% and 20% of the maximal Lorentz factor. This fraction is plotted as a function of the number of zones in the x-direction of the two-dimensional square mesh. Fig. 3a shows the result for a rotor problem with w=9.95, corresponding to a Lorentz factor of 10. Fig. 3b shows the result for a rotor problem with w=9.9944, corresponding to a Lorentz factor of 30.*

*Figs. 4a, 4b, 4c and 4d show the density, gas pressure, Lorentz factor and magnetic field strength for the RMHD rotor problem with a starting Lorentz factor of 10. The simulation was run on a 2500×2500 zone mesh with a third order scheme and stopped at a time of 0.4.*

*Figs. 5a, 5b, 5c and 5d show the density, gas pressure, Lorentz factor and magnetic field strength for the RMHD rotor problem with a starting Lorentz factor of 30. The simulation was run on a 4700×4700 zone mesh with a third order scheme and stopped at a time of 0.4.*

*Figs. 6a, 6b, 6c and 6d show the $\log_{10}$ density, $\log_{10}$ gas pressure, Lorentz factor and magnetic field strength for the RMHD blast problem The simulation was run on a 400×400 zone mesh with a fourth order scheme and stopped at a time of 4.*

*Figs. 7a, 7b and 7c show the initial density, Lorentz factor and magnitude of the magnetic field for the RMHD shock-vortex interaction problem that was run on a 500×500 zone mesh with a third order scheme to a stopping time of 10. Fig. 7d, 7e and 7f show the same variables at a time of 3.4 when the vortex has passed halfway through the shock. Fig. 7g, 7h and 7i show the same variables at the final time. We see that the vortex has not been disrupted by its passage through the shock.*

*Figs. 8a, 8b and 8c show the $\log_{10}$ density, $\log_{10}$ of magnitude of the magnetic field and Lorentz factor for an RMHD jet with internal Mach number of 6 and an initial Lorentz factor of 7. A second order scheme was run on a 120×500 zone cylindrical mesh to a time of 126.*

*Figs. 9a and 9b show the scaled maximum z-velocity and maximum z-component of the magnetic field as a function of time for the RMHD Alfven wave decay problem. The simulations were run with second, third and fourth order accurate ADER-WENO schemes. The dissipation decreases with increasing order till it becomes almost imperceptible at fourth order.*

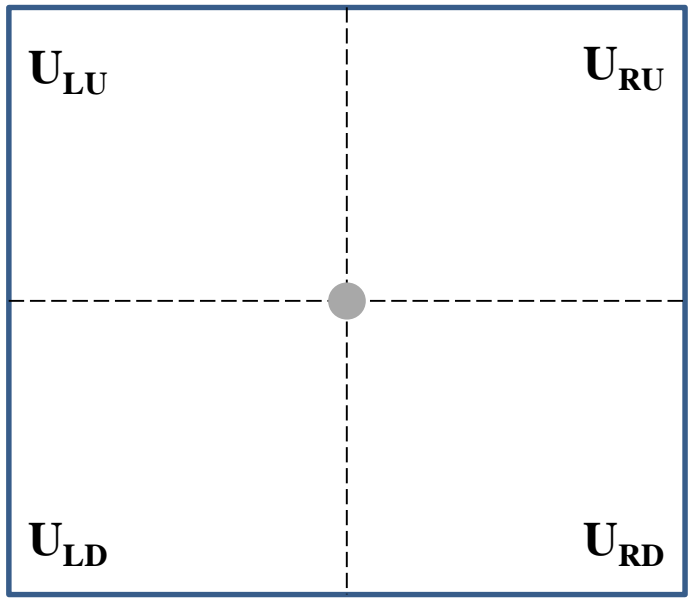 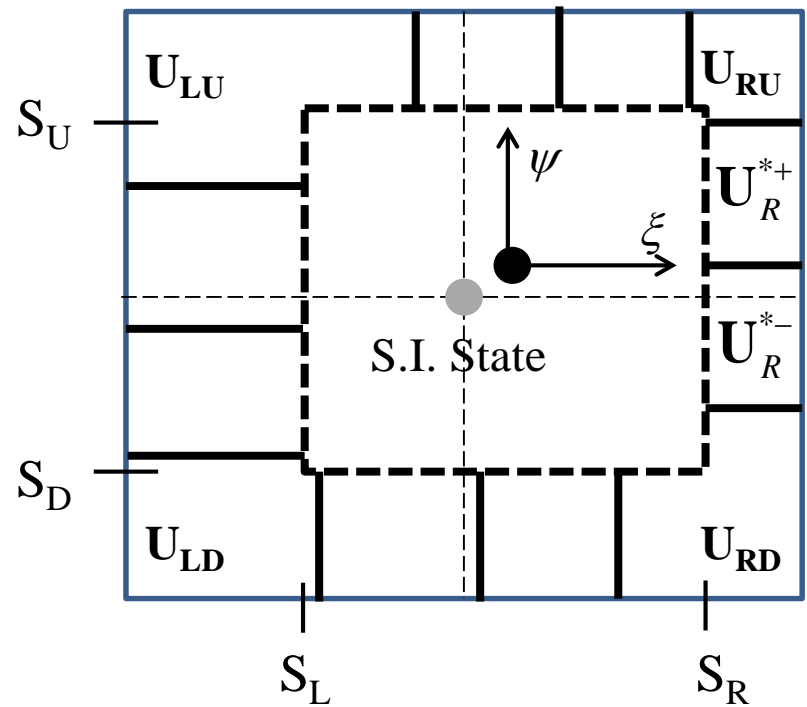

Fig 1a shows the initial situation where the four states come together at a vertex (shown by the gray dot). In time, the states evolve and we schematically show the wave model of an HLLC Riemann solver with two extremal waves and one central wave. In Fig. 1b we show four one-dimensional Riemann problems which interact with one-another to produce a strongly interacting state (S.I. State); see the thick dashed lines. The extremal speeds associated with the S.I. State are also shown. In Fig. 1b, the strongly interacting state itself provides the resolved state at the vertex.

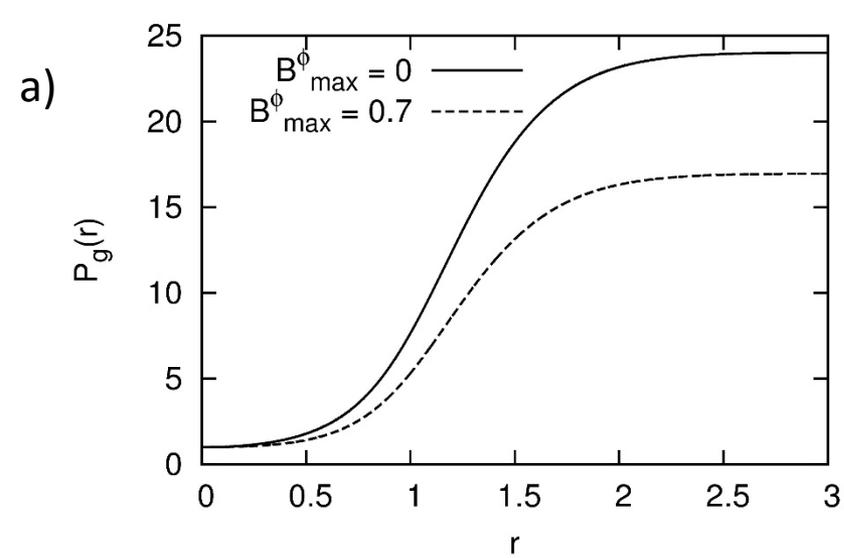
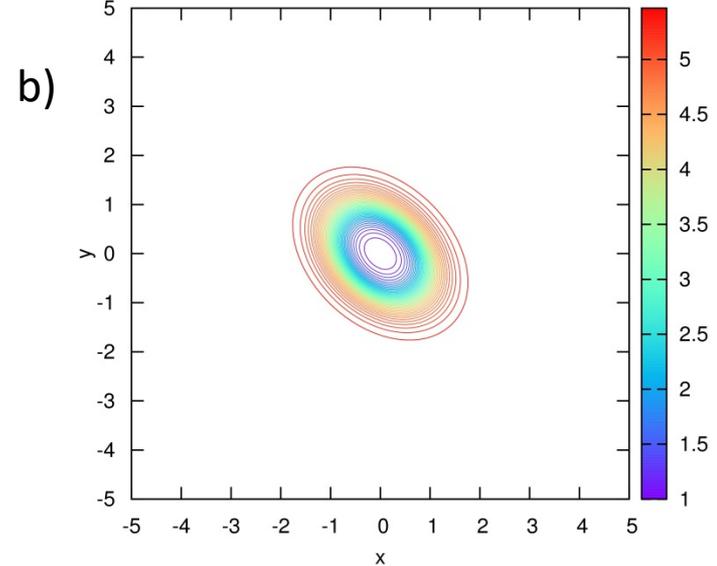
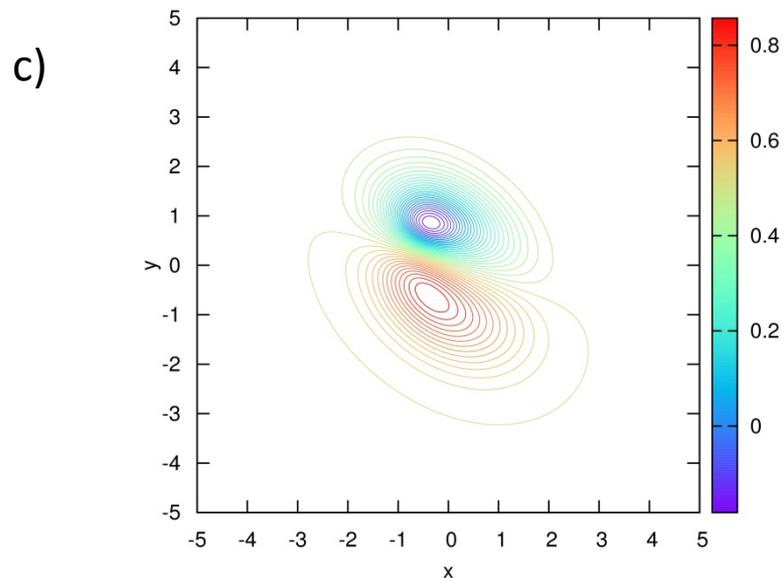
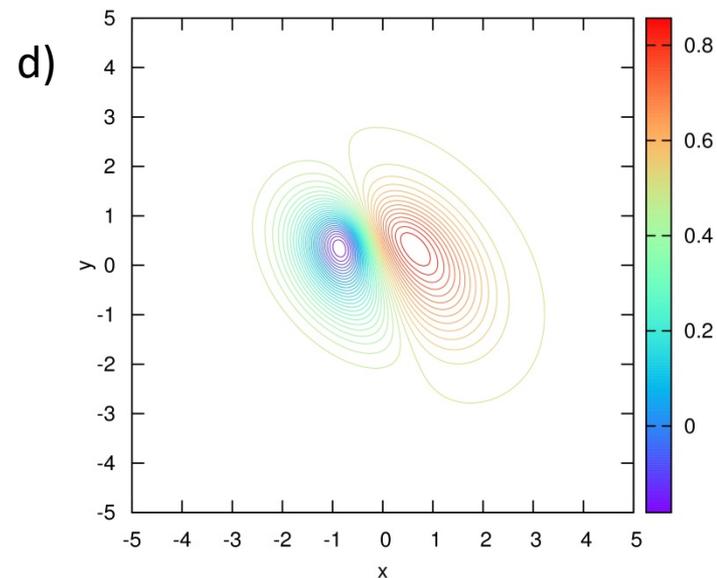

*Fig. 2a shows the run of thermal pressure as a function of radius for the vortices used here in the relativistic hydrodynamics and RMHD cases. Fig. 2b we show the density profile of the RMHD vortex on the computational mesh at the initial time. Figs. 2c and 2d show the x-velocity and y-velocity respectively.*

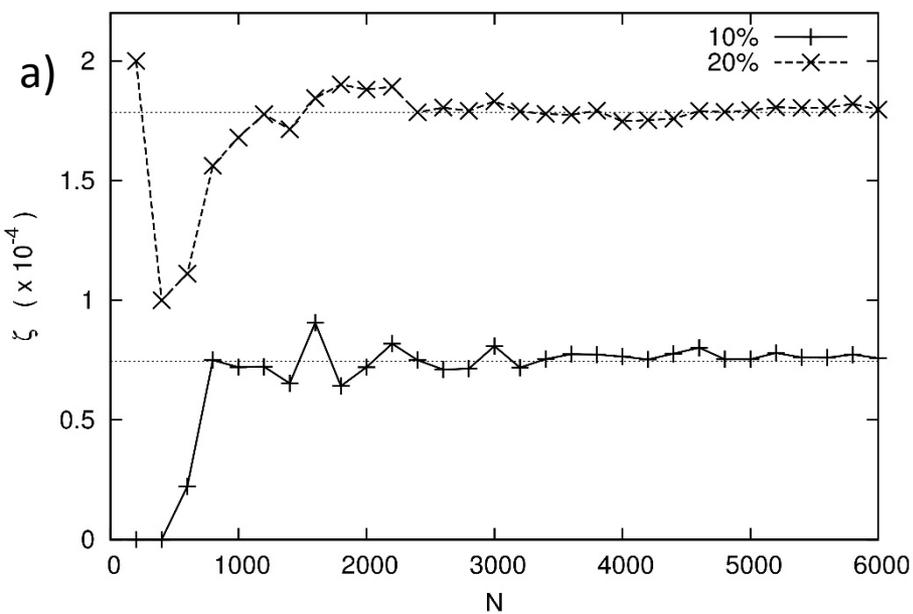 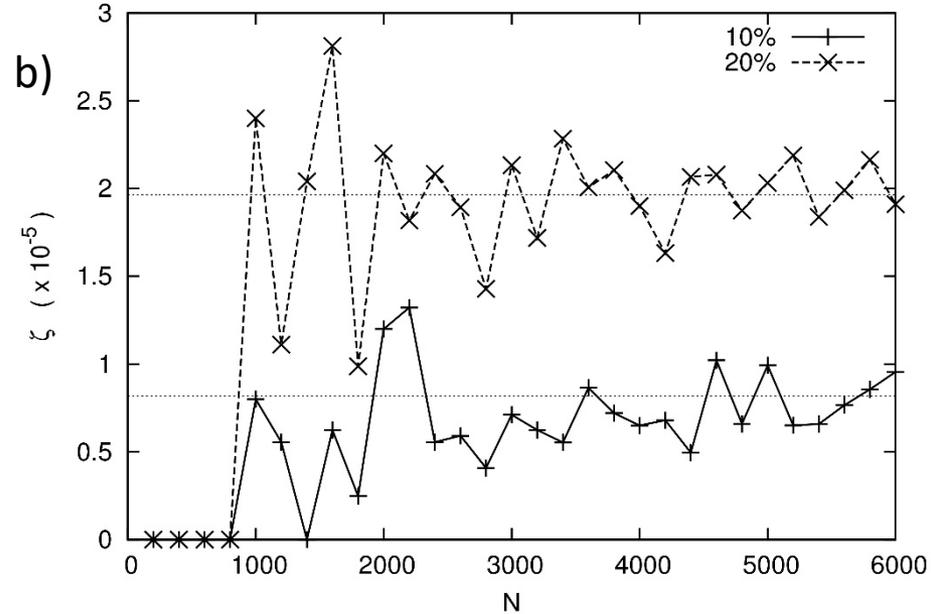

*Fig. 3 plots the fraction of zones on the mesh for the rotor problem that are within 10% and 20% of the maximal Lorentz factor. This fraction is plotted as a function of the number of zones in the x-direction of the two-dimensional square mesh. Fig. 3a shows the result for a rotor problem with w=9.95, corresponding to a Lorentz factor of 10. Fig. 3b shows the result for a rotor problem with w=9.9944, corresponding to a Lorentz factor of 30.*

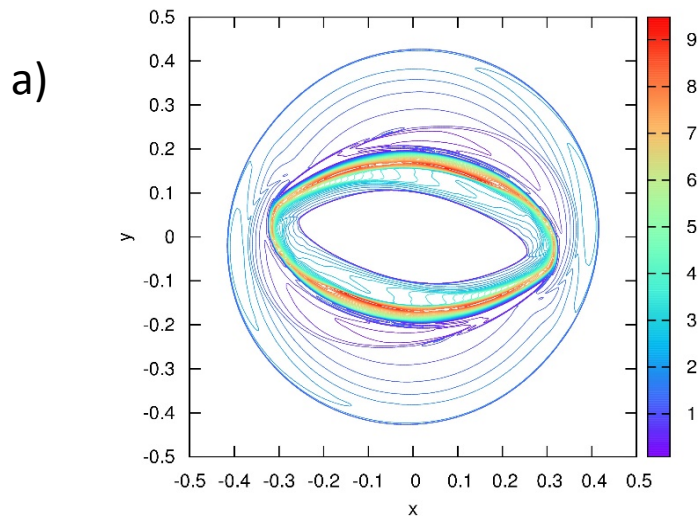 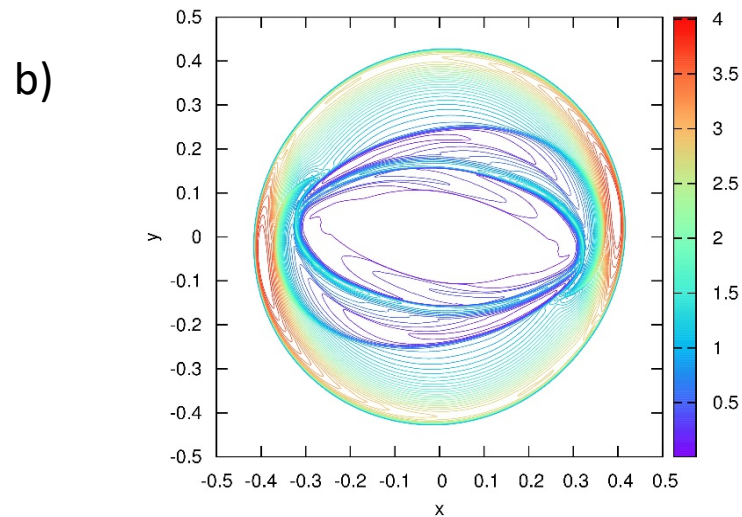
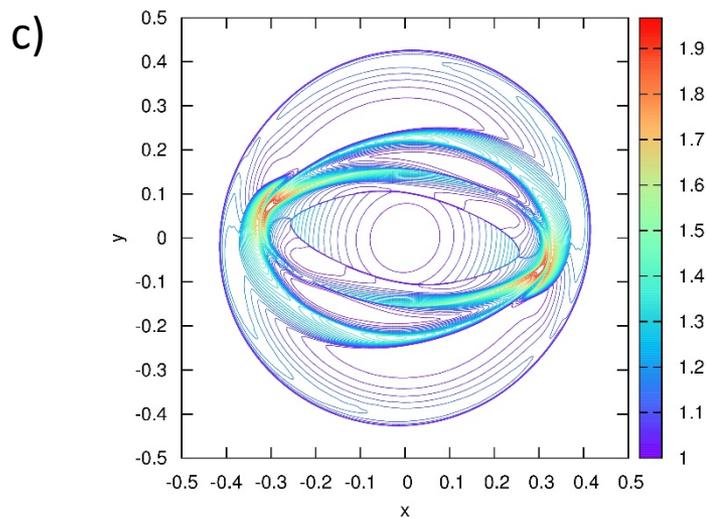 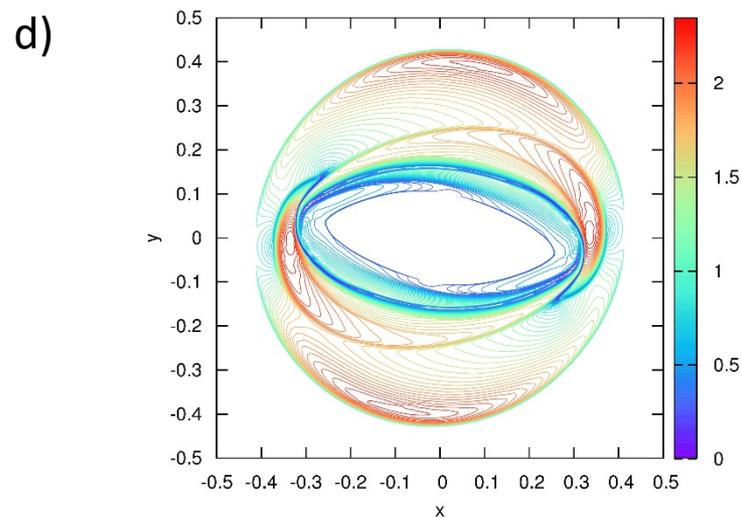

*Figs. 4a, 4b, 4c and 4d show the density, gas pressure, Lorentz factor and magnetic field strength for the RMHD rotor problem with a starting Lorentz factor of 10. The simulation was run on a 2500×2500 zone mesh with a third order scheme and stopped at a time of 0.4.*

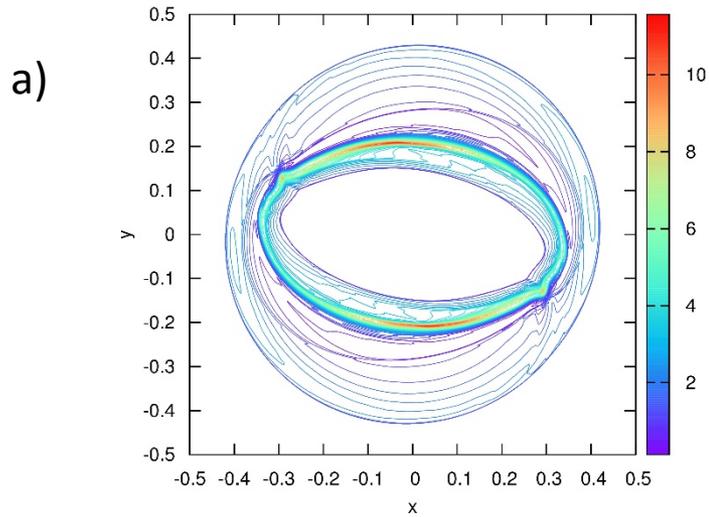 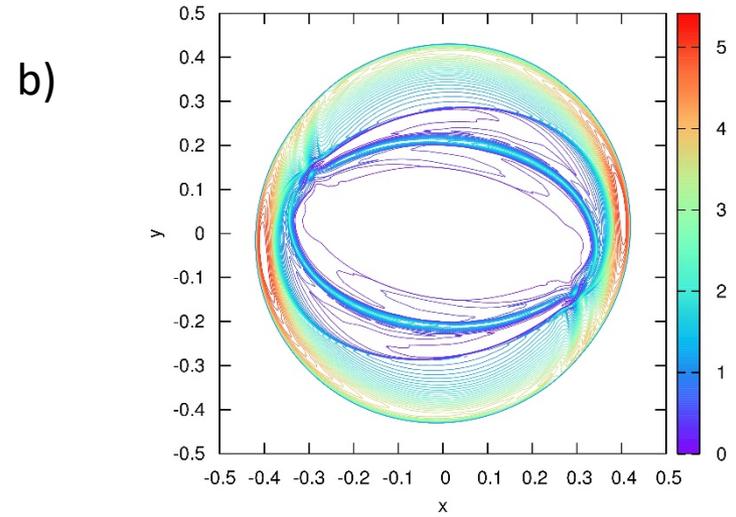
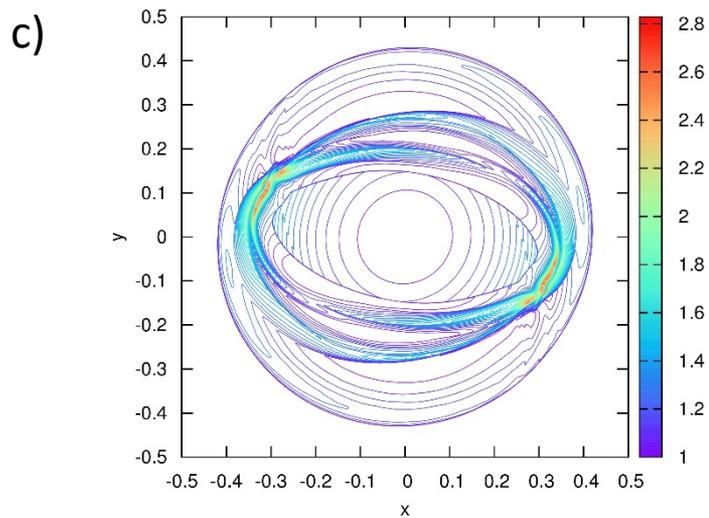 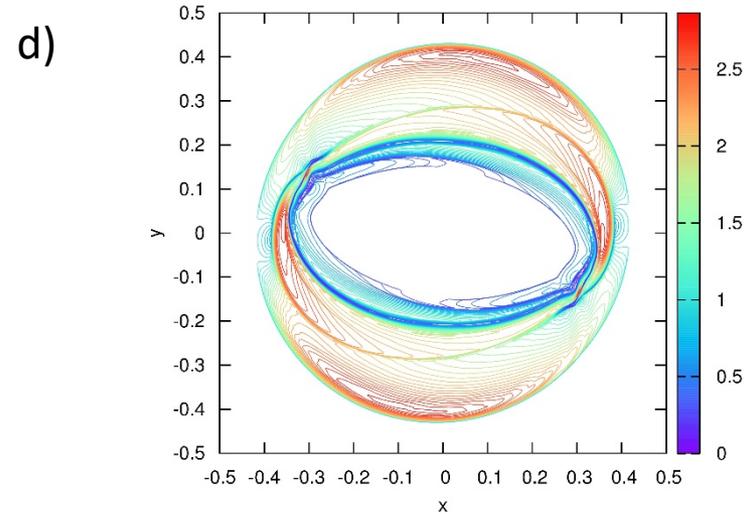

*Figs. 5a, 5b, 5c and 5d show the density, gas pressure, Lorentz factor and magnetic field strength for the RMHD rotor problem with a starting Lorentz factor of 30. The simulation was run on a 4700×4700 zone mesh with a third order scheme and stopped at a time of 0.4.*

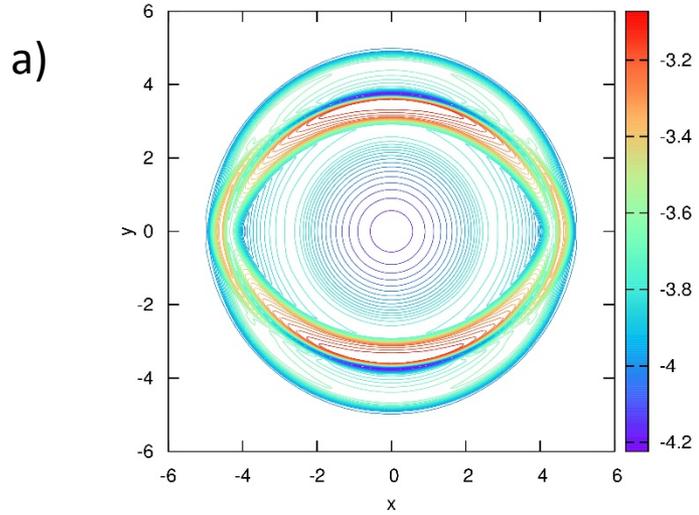 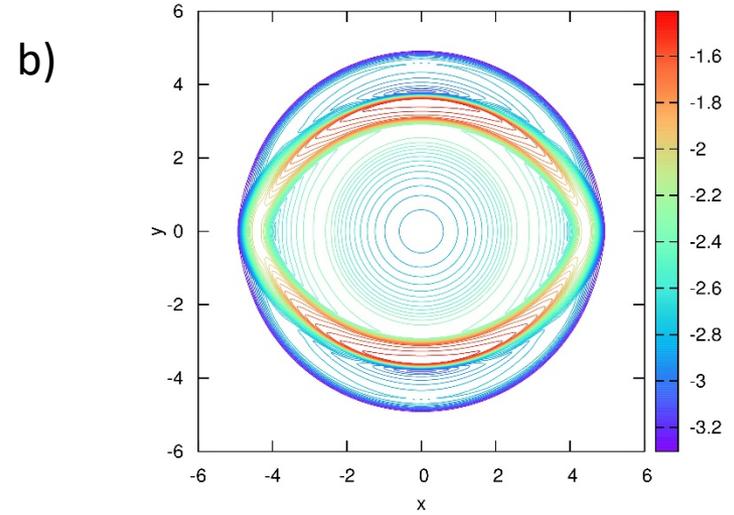
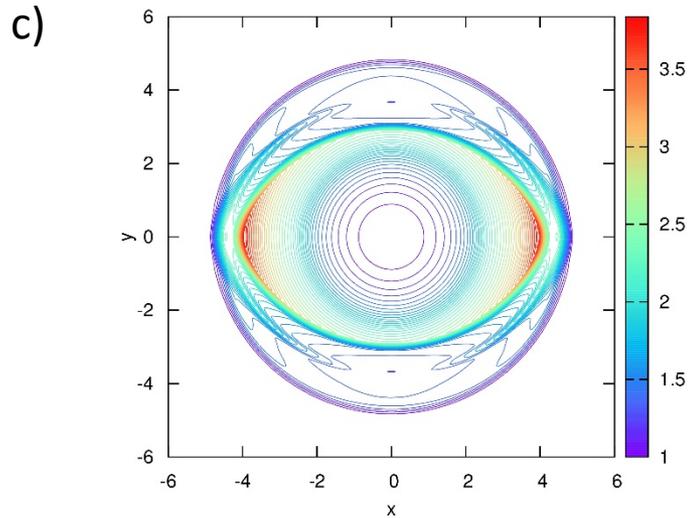 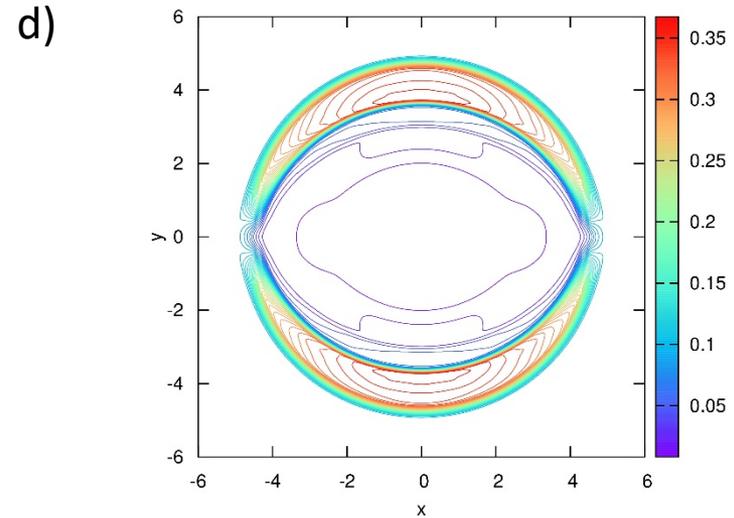

*Figs. 6a, 6b, 6c and 6d show the $\log_{10}$ density, $\log_{10}$ gas pressure, Lorentz factor and magnetic field strength for the RMHD blast problem The simulation was run on a 400×400 zone mesh with a fourth order scheme and stopped at a time of 4.*

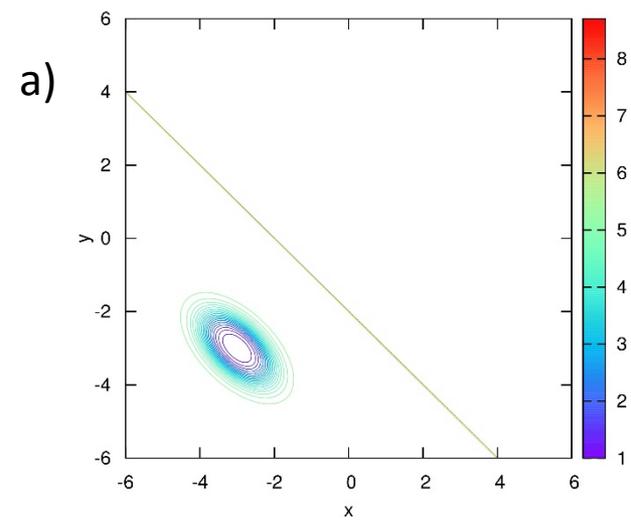 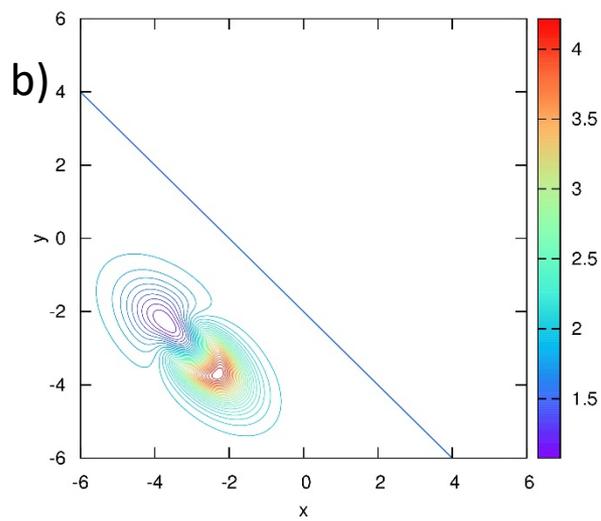 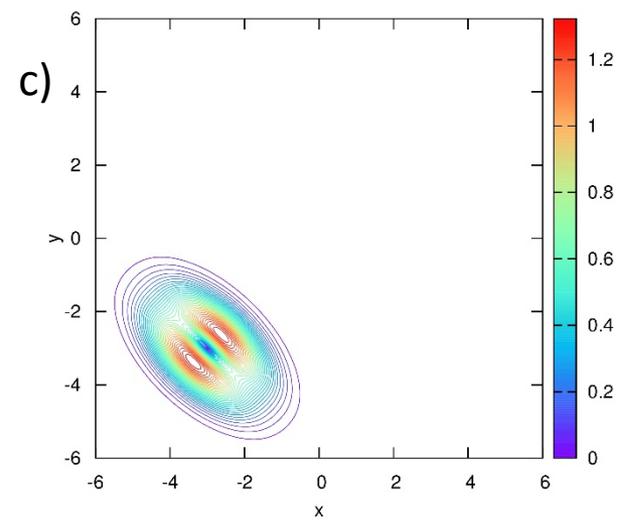
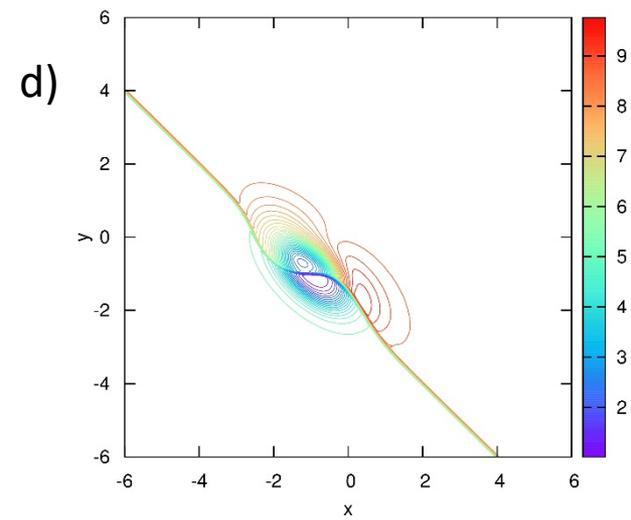 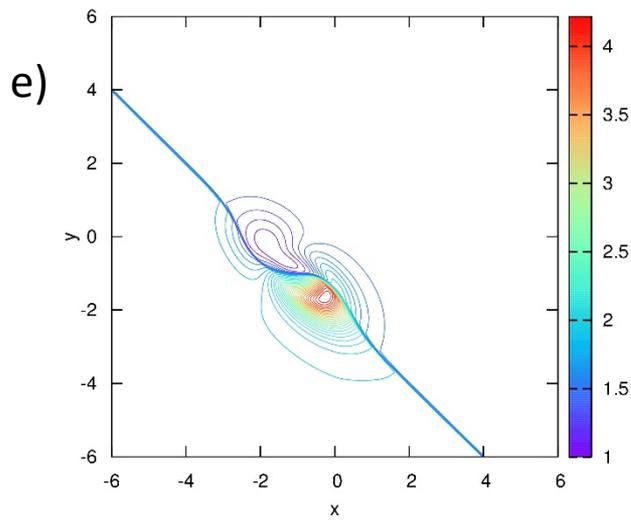 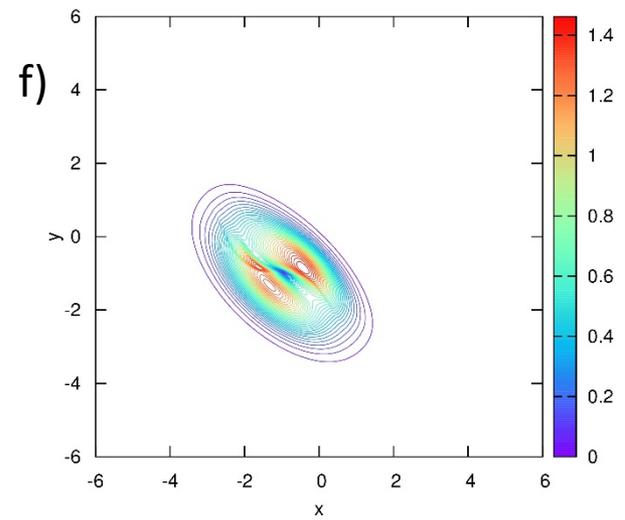

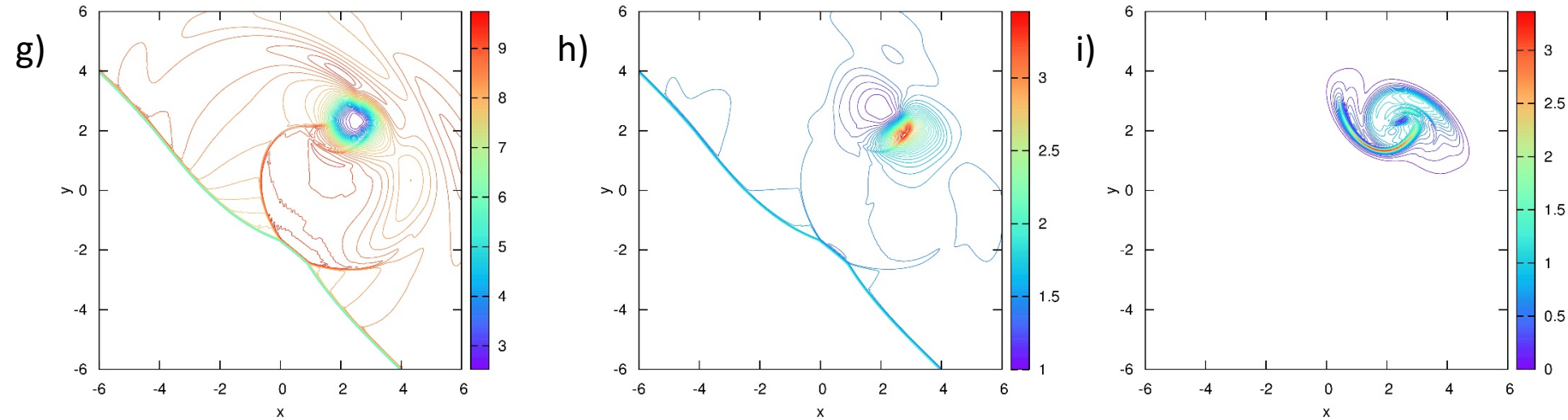

Figs. 7a, 7b and 7c show the initial density, Lorentz factor and magnitude of the magnetic field for the RMHD shock-vortex interaction problem that was run on a 500×500 zone mesh with a third order scheme to a stopping time of 10. Fig. 7d, 7e and 7f show the same variables at a time of 3.4 when the vortex has passed halfway through the shock. Fig. 7g, 7h and 7i show the same variables at the final time. We see that the vortex has not been disrupted by its passage through the shock.

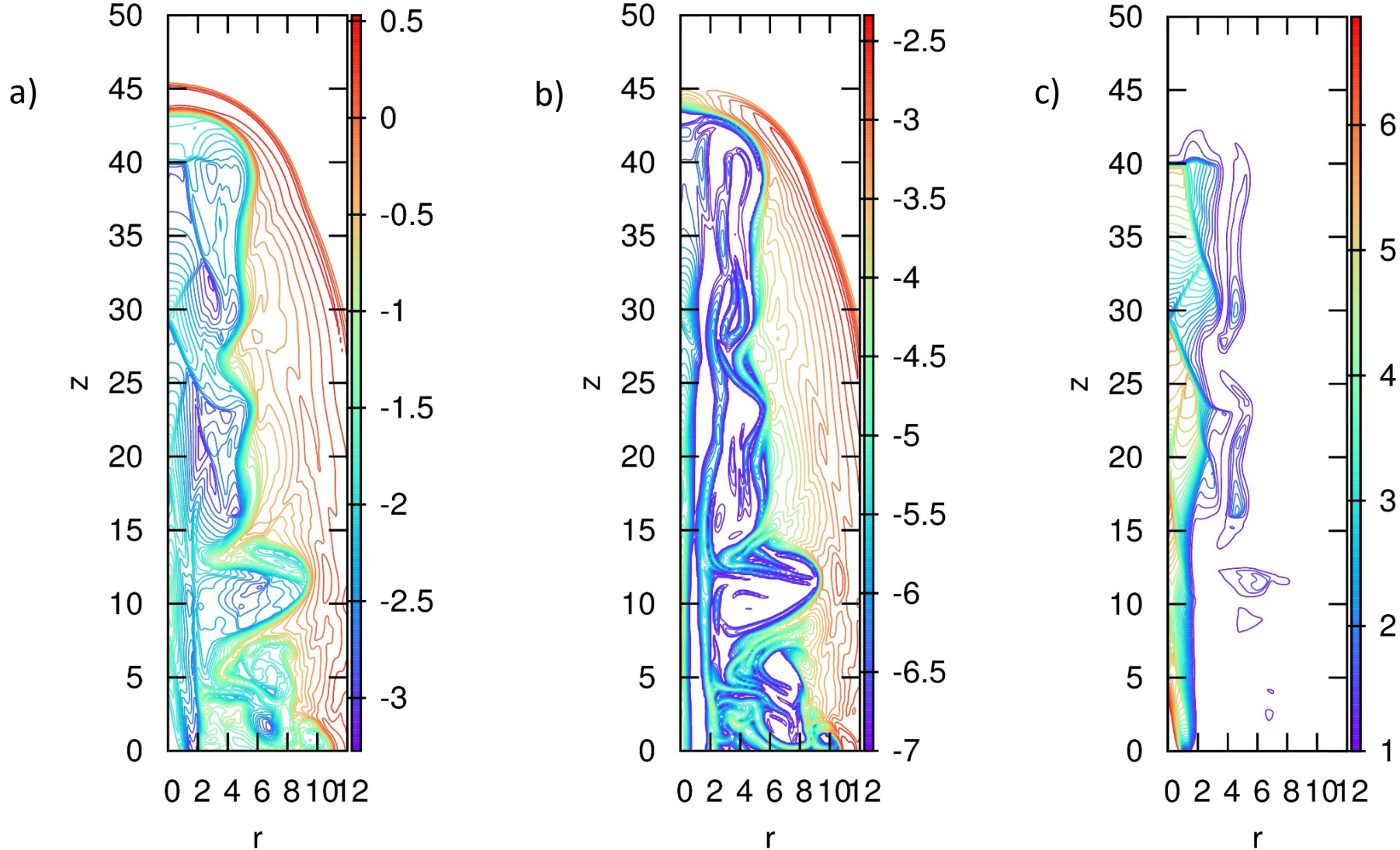

*Figs. 8a, 8b and 8c show the $\log_{10}$ density, $\log_{10}$ of magnitude of the magnetic field and Lorentz factor for an RMHD jet with internal Mach number of 6 and an initial Lorentz factor of 7. A second order scheme was run on a 120×500 zone cylindrical mesh to a time of 126.*

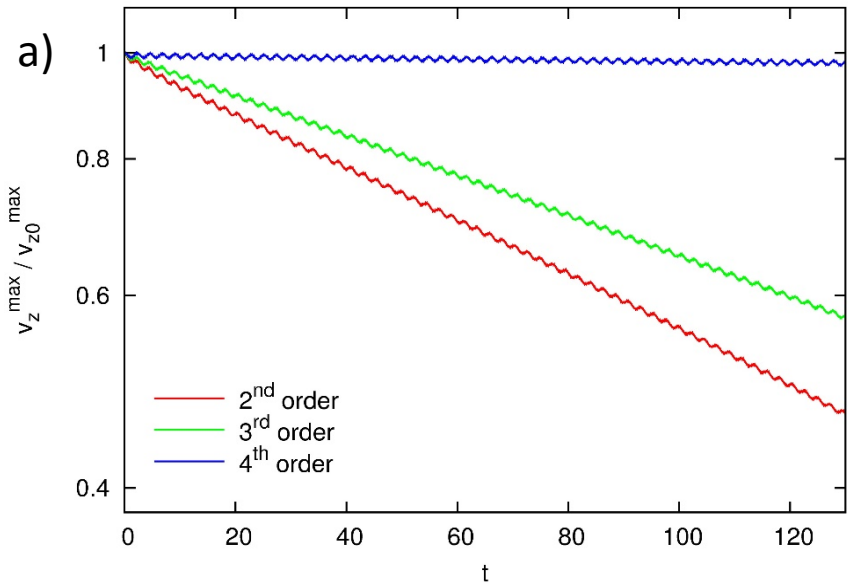 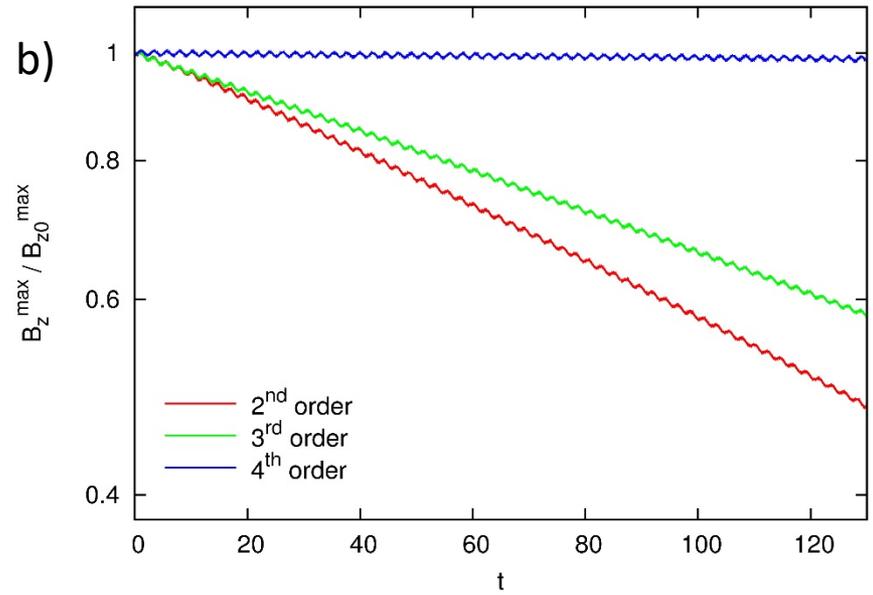

Figs. 9a and 9b show the scaled maximum z-velocity and maximum z-component of the magnetic field as a function of time for the RMHD Alfven wave decay problem. The simulations were run with second, third and fourth order accurate ADER-WENO schemes. The dissipation decreases with increasing order till it becomes almost imperceptible at fourth order.